\documentclass{article}
\usepackage{amsmath,epsfig,amsfonts,amsthm,amstext}
\usepackage{amssymb}
\usepackage{subfig}
\usepackage{graphicx}
\usepackage{sidecap}
\usepackage[
  unicode=true,
  bookmarks=false,
  pdfpagelabels=false,
  hyperfootnotes=false,
  hyperindex=false,
  pageanchor=false,
  colorlinks=false,
]{hyperref}
\usepackage{cleveref}
\usepackage[font={small,it}]{caption}
\usepackage[affil-it]{authblk}

\title{Geodesic Transport Barriers in Jupiter's Atmosphere:\\
 A Video-Based Analysis }

\author{Alireza Hadjighasem\thanks{alirezah@ethz.ch}}
\author{George Haller\thanks{Email address for correspondence: georgehaller@ethz.ch}}
\affil{Institute of Mechanical Systems, Department of Mechanical and Process Engineering, ETH Z\"{u}rich, Leonhardstrasse 21, 8092 Z\"{u}rich, Switzerland}

\begin{document}
\maketitle
\begin{abstract}
Jupiter's zonal jets and Great Red Spot are well known from still
images. Yet the planet's atmosphere is highly unsteady, which suggests
that the actual material transport barriers delineating its main features
should be time-dependent. Rare video footages of Jupiter's clouds
provide an opportunity to verify this expectation from optically reconstructed
velocity fields. Available videos, however, provide short-time and
temporally aperiodic velocity fields that defy classical dynamical
systems analyses focused on asymptotic features. To this end, we use
here the recent theory of geodesic transport barriers to uncover finite-time
mixing barriers in the wind field extracted from a video captured
by NASA's Cassini space mission. More broadly, the approach described
here provides a systematic and frame-invariant way to extract dynamic
coherent structures from time-resolved remote observations of unsteady
continua. 
\end{abstract}

\section{Introduction}

Jupiter's size is 1,300 times that of the Earth. Its mass is more
than twice the mass of all planets in our solar system combined. Jupiter's
fast rotation --- one in 10 hours --- creates strong jet streams that
smear its clouds into bands of zones and belts of almost constant
latitude. Another frequently discussed feature of Jupiter is its Great
Red Spot (GRS), the largest and longest-living known atmospheric vortex.
The GRS is a nearly two-dimensional feature that is apparently unrelated
to the topography of the planet \cite{Golitsyn}. Such vortices abound
in nature, but GRS's size, long-term persistence, and temporal longitudinal
oscillations make it unique.

Jupiter's atmospheric features are generally inferred from still images,
but should clearly be time-dependent objects in the planet's turbulent
atmosphere. In fluid dynamics, evolving features in a complex flow
are often referred to as transport barriers, which in turn are described
as objects that cannot be crossed by other fluid trajectories. However
intuitive this characterization of barriers might sound, it actually
labels all points in a moving continuum as part of a barrier. This
is because any material surface (i..e, a connected, time-evolving
set of fluid trajectories) is impenetrable to other trajectories in
a flow with unique trajectories. Indeed, a transport barrier is more
than just an impenetrable material object: it is a material surface
that remains coherent by withstanding stretching and filamentation
\cite{Haller14}.

A recent approach in dynamical systems theory seeks transport barriers
in unsteady flows as key material surfaces with exceptionally coherent
features in their deformation. These exceptional material surfaces,
Lagrangian coherent structures (or LCSs), were initially defined as
most attracting, repelling or shearing material surfaces \cite{HallerYuan00}.
Turning this definition into computable mathematical results has proven
challenging, prompting instead a widespread use of intuitive diagnostics
in LCS detection (see \cite{Peacock10} and \cite{Shadden12} for
reviews). Alternative approaches have also been developed in the meantime to target regions enclosed by LCSs (see \cite{Froyland10,Mezic10,Ma14}).

More recent advances have re-addressed the unsolved mathematical challenges
by seeking an LCS as a stationary curve of the Lagrangian strain or
shear functional computed along material lines \cite{black_hole,faraz13}.
These variational methods (here collectively referred to as geodesic
LCS theory) uncover LCSs as null-geodesics of appropriate strain tensor
fields computed from the deformation field. In contrast to the visual
assessment of features in intuitive diagnostic fields, geodesic LCS
theory renders transport barriers as smooth, parametrized curves that
are exact solutions of well-defined stationarity principles. These
solutions depend only on frame-invariant tensor fields, and hence
remain the same in translating and rotating frames. Given these advantages,
we use geodesic LCS detection in the present work to uncover unsteady
transport barriers in Jupiter's atmosphere.

Locating geodesic LCSs requires a time-resolved
velocity field. For Jupiter, a representative two-dimensional
wind-velocity field can be obtained via image-correlation analysis
of available cloud videos. In this work, we apply the Advection Corrected
Correlation Image Velocimetry (ACCIV) method \cite{ACCIV} to obtain
a high-density, time-resolved representation of Jupiter's wind field
from an enhanced version of a video taken by the Cassini mission of
NASA in 2000.

Our main objective in this paper is twofold. First, we would like
to provide a technical review of geodesic LCS theory, summarizing
various aspects of the theory from different sources in a unified
notation. Second, we wish to show how this theory reveals details
of objectively (i.e., frame-invariantly) defined coherent structures
in an unsteady flow known only from remote optical sensing. This flow,
the wind field of Jupiter reconstructed from the Cassini video, embodies
all the major challenges to practical transport barrier detection.
First, the data covers a relatively short time period; second, it
is temporally aperiodic; and third, it was captured in a non-inertial
frame of reference. These complications necessitate the correct handling
of finite-time (as opposed to asymptotic) dynamical systems structures;
an abandonment of recurrence and temporal convergence assumptions;
and the use of objective (i.e., frame-invariant) methods (cf. \cite{Haller14}
for details on these features).

In mathematical terms, our analysis uncovers elliptic and parabolic invariant manifolds in a non-autonomous, temporally aperiodic, finite-time dynamical system. Until recently, a precise definition and extraction of such manifolds has been an unsolved problem even for analytically defined velocity models. As we discuss in detail below, these newly identified dynamical structures support earlier physics-based conclusions obtained by others for Jupiter's atmosphere. Specifically, we confirm model-based transport predictions by Beron-Vera et al. \cite{Beron_Vera08b} for zonal jet cores, and find consistency with a geometric circulation model around the GRS proposed by Conrath et al. \cite{Conrath81} and de Pater et al. \cite{Pater10}. In addition, we uncover the Lagrangian signature of chevron-type atmospheric features discovered recently in Jupiter's atmosphere by Simon-Miller et al. \cite{Simon12}.

\section{Set-up}

Consider a two-dimensional unsteady velocity field 
\begin{equation}
\dot{x}=v(x,t),\quad x\in U\subset\mathbb{R}^{2},\quad t\in[t_{0},t_{1}],\label{eq:dynsys}
\end{equation}
whose trajectories $x(t;t,x_{0})$ define a finite-time flow map $F_{t_{0}}^{t}(x_{0}):x_{0}\mapsto x(t;t,x_{0})$
for times $t\in[t_{0},t_{1}]$ over the spatial domain $U$. A \emph{material
line} $l(t)$ is a smooth curve of initial conditions under the flow,
satisfying 
\begin{equation}
l(t)=F_{t_{0}}^{t}\left(l(t_{0})\right).\label{eq:matline}
\end{equation}
Any material line spans a two-dimensional invariant manifold in the
three-dimensional extended phase space of the $(x,t)$ variables.
Lagrangian Coherent Structures (LCSs) in two-dimensions can loosely
be defined as exceptional material curves that end up shaping trajectory
patterns. This definition will be made more precise in the next section.
Here we only observe that an LCS, just as any material line, is an
invariant manifold in the extended phase space $U\times[t_{0},t_{1}]$,
but generally not in the phase space $U$.

To asses the influence of specific material lines on trajectories,
we will need a classic measure of flow deformation, the right Cauchy--Green
strain tensor 
\begin{equation}
C_{t_{0}}^{t}(x_{0})=({\nabla F_{t_{0}}^{t}})^{\intercal}\nabla F_{t_{0}}^{t},\label{eq:CG}
\end{equation}
where $\nabla F_{t_{0}}^{t}$ denotes the gradient of the flow map,
and the symbol $\intercal$ indicates matrix transposition. The tensor
$C_{t_{0}}^{t}$ is symmetric and positive definite; it has two
positive eigenvalues $0<\lambda_{1}\leq\lambda_{2}$ and an orthonormal
eigenbasis $\left\{ \xi_{1},\xi_{2}\right\} $ satisfying 
\begin{align}
 & C_{t_{0}}^{t}(x_{0})\xi_{i}(x_{0})=\lambda_{i}(x_{0})\mathbf{\xi}_{i}(x_{0}),\quad\left|\xi_{i}(x_{0})\right|=1,\quad i\in\{1,2\},\nonumber \\
 & \xi_{2}(x_{0})=\Omega\xi_{1}(x_{0}),\quad\Omega=\left(\begin{array}{cc}
0 & -1\\
1 & 0
\end{array}\right).
\end{align}
The Cauchy--Green strain tensor is objective in the sense of continuum
mechanics: its invariants remain unchanged in rotating and translating
frames \cite{Gurtin81}. We will also need to use the symmetric part
of the tensor $C_{t_{0}}^{t}(x_{0})\Omega$, defined as 
\begin{equation}
D_{t_{0}}^{t}(x_{0})=\frac{1}{2}\left[C_{t_{0}}^{t}(x_{0})\Omega-\Omega C_{t_{0}}^{t}(x_{0})\right].
\end{equation}

\section{Geodesic LCS theory}\label{sec:LCS}

A general material line of system \eqref{eq:dynsys} experiences shear
and strain in its deformation. Both shear and strain depend continuously
on initial conditions owing to the continuity of the map $F_{t_{0}}^{t}$.
The averaged strain and shear within a strip of $\epsilon$-close
material lines, therefore, generically vary by an $\mathcal{O\left(\epsilon\right)}$
amount within the strip.

The geodesic theory of Lagrangian Coherent Structures (LCSs) seeks
exceptionally coherent locations where this general trend breaks down
\cite{Haller14}. Specifically, the theory searches for LCSs
as special material lines around which $\mathcal{O\left(\epsilon\right)}$
material belts show no $\mathcal{O\left(\epsilon\right)}$ variation
either in the material shear or in the material strain, both accumulated
over $[t_{0},t]$ and averaged over material lines.

These variational principles identify the time $t_{0}$ positions
of LCSs as stationary curves of the material-line-averaged Lagrangian
shear or Lagrangian strain functionals. Both principles reveal that
the initial positions of shearless (hyperbolic and parabolic) and
strainless (elliptic) LCSs are null-geodesics of appropriate
tensor fields \cite{black_hole,faraz13}. Later
positions of these LCSs can be found by advecting the null-geodesics
under the flow map, as described in \eqref{eq:matline}. Recent results
\cite{Karrasch14} eliminate numerical instabilities arising in the
advection of hyperbolic LCSs. For the elliptic and parabolic LCS considered
here, the advection process \eqref{eq:matline} is stable.

We summarize below the main results from \cite{faraz13} for parabolic
LCSs (or generalized jet cores) and from \cite{black_hole} for elliptic
LCSs (or generalized KAM curves). Parabolic LCSs are expected to identify
the unsteady cores of Jupiter's zonal jets. The largest member
of a nested family of elliptic LCSs is expected to mark the Lagrangian
boundary of the Great Red Spot. The differential equations rendering
these geodesic LCSs only depend on the invariants of the tensor field
$C_{t_{0}}^{t}(x_{0})$, and hence give frame-invariant results. This
objectivity of geodesic LCSs is especially important when the underlying
velocity field \eqref{eq:dynsys} is reconstructed in a moving frame,
such as the frame of the Cassini spacecraft flying by Jupiter.

\subsection{Parabolic LCSs\label{sub:Parabolic-LCSs}}

\begin{figure}
\centering 
\includegraphics[width=0.8\textwidth]{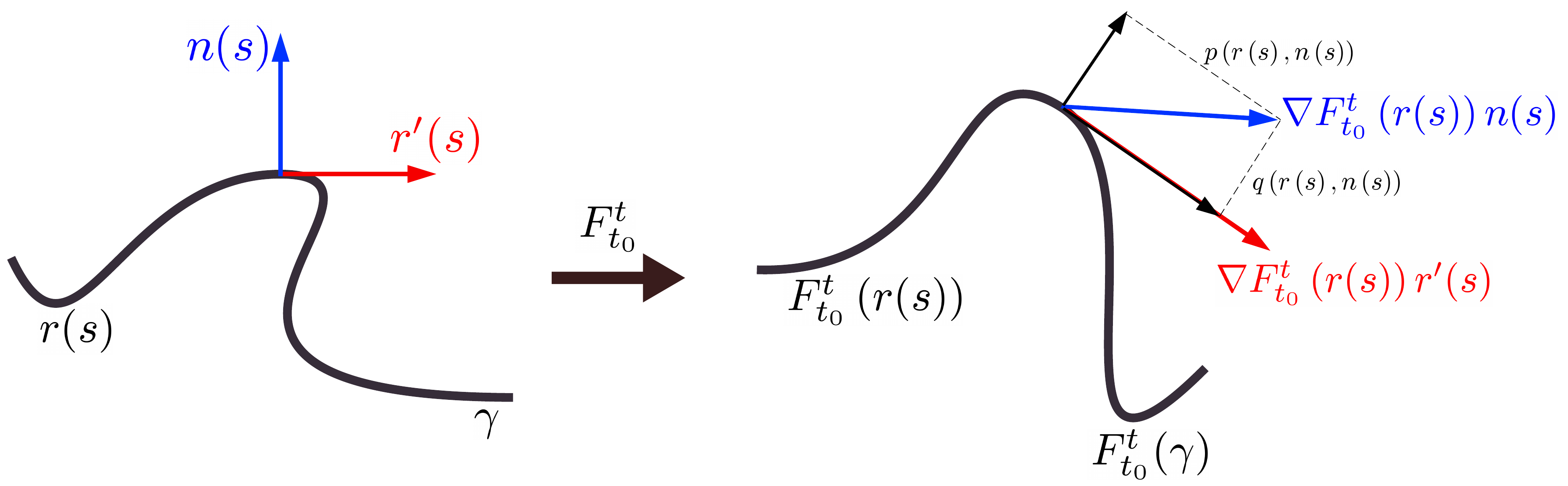}
\caption{The evolution of material line $\gamma$ as well as its tangent and
normal vectors under the linearized flow map $DF_{t_{0}}^{t}$.}
\label{fig:curve_deformation} 
\end{figure}

We consider an initial material line $\gamma:=l(t_{0})$, parametrized
as $r(s)$ with $s\in[s_{1},s_{2}]$. The tangent vectors along $\gamma$
are then given by $r'(s)$, and a smoothly varying unit normal along
$\gamma$ is given by $n(s)=\Omega r'(s)/\left|r'(s)\right|$. The
flow map $F_{t_{0}}^{t}$ maps $\gamma$ to its time $t$ position,
as shown in Fig. \ref{fig:curve_deformation}. As in \cite{faraz13},
we define the Lagrangian shear $p(r(s),n(s))$ along $\gamma$ as
the tangential projection of the linearly advected normal vector $\nabla F_{t_{0}}^{t}(r(s))n(s)$.

The averaged Lagrangian shear experienced by $\gamma$ over the time interval
$\left[t_{0},t\right]$ is given by \cite{faraz13}
\[P(\gamma)=\intop_{\gamma}p(r(s),r'(s))ds=\intop_{\gamma}\sqrt{\frac{\left\langle r^{\prime}(s),D_{t_{0}}^{t}(r(s))r^{\prime}(s)\right\rangle }{\left\langle r^{\prime}(s),C_{t_{0}}^{t}(r(s))r^{\prime}(s)\right\rangle \left\langle r^{\prime}(s),r^{\prime}(s)\right\rangle }}ds.\]
We seek shearless LCSs as material lines with no leading
order variation in their averaged Lagrangian shear. On the time $t_{0}$
position of such LCSs, the first variation of $P$ must necessarily
vanish, i.e., 
\begin{equation}
\delta P(\gamma)=0.
\label{eq:shearlessvari}
\end{equation}

\begin{figure}
\centering 
\subfloat[\label{fig:strain_stretch_chain}]{\includegraphics[width=0.7\textwidth]{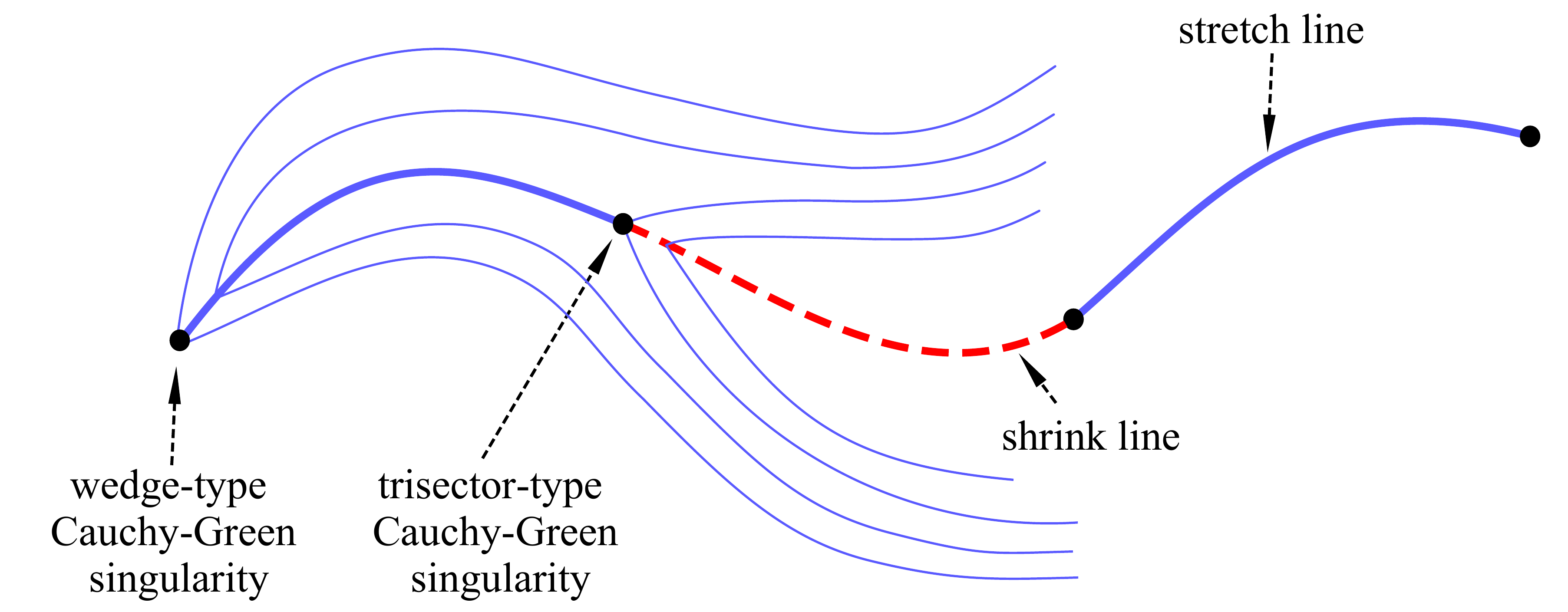}}

\subfloat[\label{fig:singularity_topology}]{\includegraphics[width=0.4\textwidth]{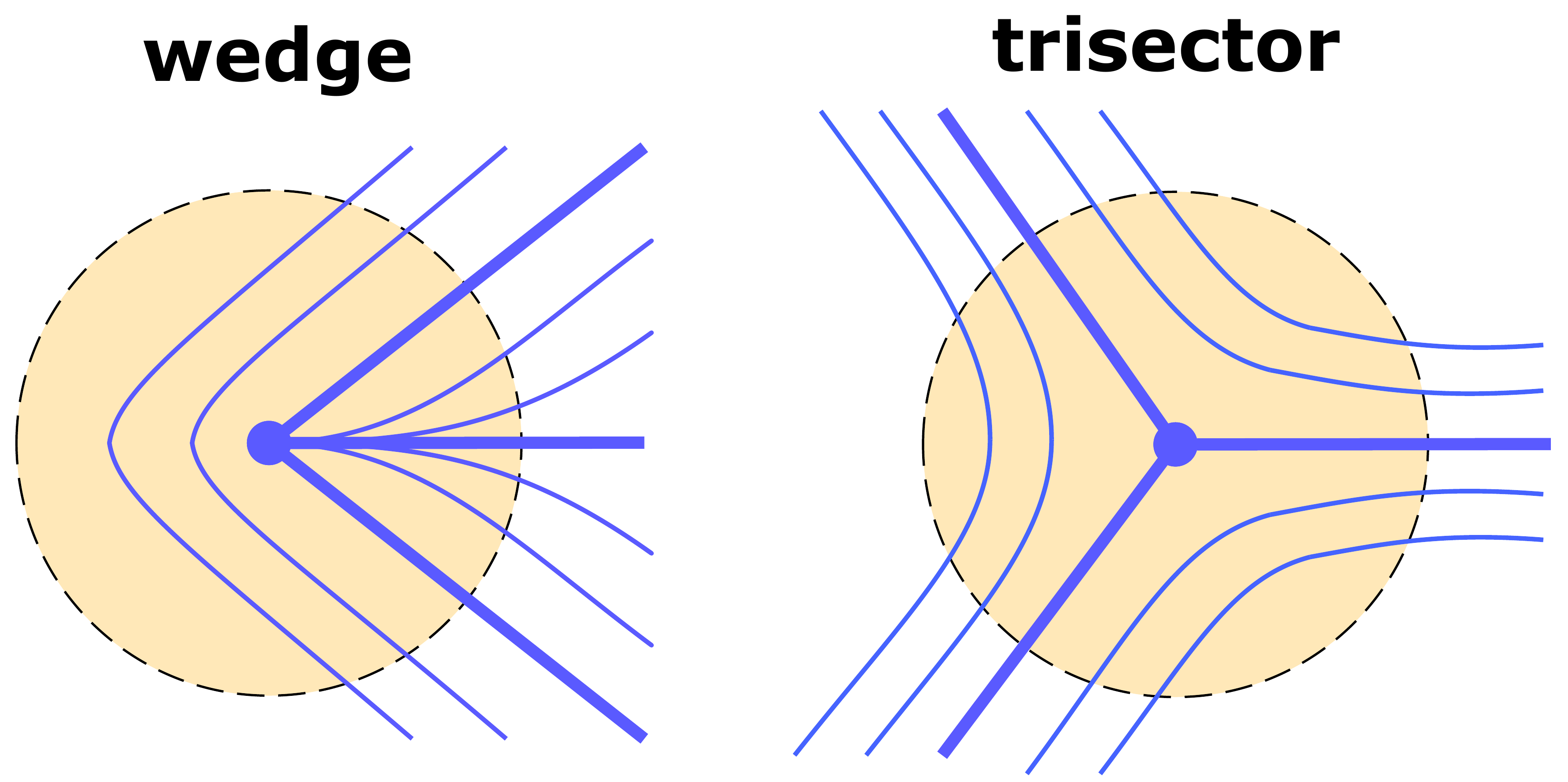}}

\caption{(a) A parabolic LCS is a structurally stable, alternating chain of shrink-stretch curve segments that connect Cauchy--Green singularities. A single shrink line (red) in such a chain is superimposed on the local stretch-line geometry (blue) near the LCS. (b) Topology of stretch lines around wedge-type and trisector-type Cauchy--Green singularities.}

\label{fig:strain_stretch} 
\end{figure}

The most readily observed solutions of \eqref{eq:shearlessvari} are
those obtained under the largest possible set of admissible variations,
including changes to the endpoints of $\gamma$. As shown in \cite{faraz13},
the variational problem \eqref{eq:shearlessvari} posed with free-endpoint boundary conditions
is equivalent to finding null-geodesics connecting singularities of
the Lorentzian metric 
\[
g(u,v)=\left\langle u,D_{t_{0}}^{t}(x_{0})v\right\rangle ,
\]
where $\left\langle .,.\right\rangle $ denotes the Euclidean inner
product. The singularities of $g(u,v)$ are points where $\det D_{t_{0}}^{t}(x_{0})=0$.
These are precisely the points where the Cauchy--Green tensor field
$C_{t_{0}}^{t}(x_{0})$ has repeated eigenvalues. Following the convention
in the tensor-line literature \cite{Demarcelle93}, we refer to such points as the \emph{singularities}
of the Cauchy--Green strain tensor (Fig. \ref{fig:strain_stretch}).

All null-geodesics of the metric $g$ are found to be solutions of one of the
two ODEs 
\begin{equation}
r^{\prime}(s)=\xi_{j}(r(s)),\qquad j=1,2.\label{eq:shrinkshear}
\end{equation}
We refer to trajectories of \eqref{eq:shrinkshear} with $j=1$ as
\emph{shrink lines}, as they are compressed by the flow. Similarly,
we call trajectories of \eqref{eq:shrinkshear} with $j=2$ \emph{stretch
lines}, because they are stretched by the flow. Shrink and stretch lines are special cases of \emph{tensor
lines} used in the scientific visualization literature to illustrate
features of two-dimensional tensor fields \cite{Demarcelle93}.

Smooth null-geodesics connecting metric singularities of $g(u,v)$ are, therefore,
smooth heteroclinic chains formed by shrink lines and stretch lines
among the singularities. For observable impact on mixing, we focus
on such null-geodesic chains that are structurally stable and locally
unique. This requirement restricts the shrink-stretch chains of interest
to those along which wedge- and trisector-type singularities of $C_{t_{0}}^{t}(x_{0})$
alternate (cf. Fig. \ref{fig:strain_stretch} and \cite{faraz13} for details).

The time $t_{0}$ positions of\emph{ parabolic LCS}s (Lagrangian jet
cores) are defined as tensor-line chains of the type in Fig. \ref{fig:strain_stretch}
that are also weak minimizers of a quantity that measures the closeness
of the chain to being neutrally stable under advection by $F_{t_{0}}^{t}$
\cite{faraz13}. This quantity, the \emph{neutrality }of a tensorline
at a point $r$, is defined as 
\begin{equation}
\mathcal{N}_{\xi_{j}}(r)=\left(\sqrt{\lambda_{k}(r)}-1\right)^{2},\qquad j\neq k.\label{eq:neutrality}
\end{equation}
A weak minimizer of $\mathcal{N}_{\xi_{j}}$ is then defined as a
trajectory of \eqref{eq:shrinkshear} that lies, together with the
nearest trench of $\mathcal{N}_{\xi_{j}}(r)$, in the same connected
component of the set of points defined by the relation $\mbox{\ensuremath{\left\langle \nabla^{2}\mathcal{N}_{\xi_{j}}(r)\xi_{k}(r),\xi_{k}(r)\right\rangle }}>0$,
with $j\neq k.$ (cf. \cite{faraz13} for more detail).

\begin{figure}
\centering 
\subfloat[\label{fig:singularities_v1}]{\includegraphics[width=0.3\textwidth]{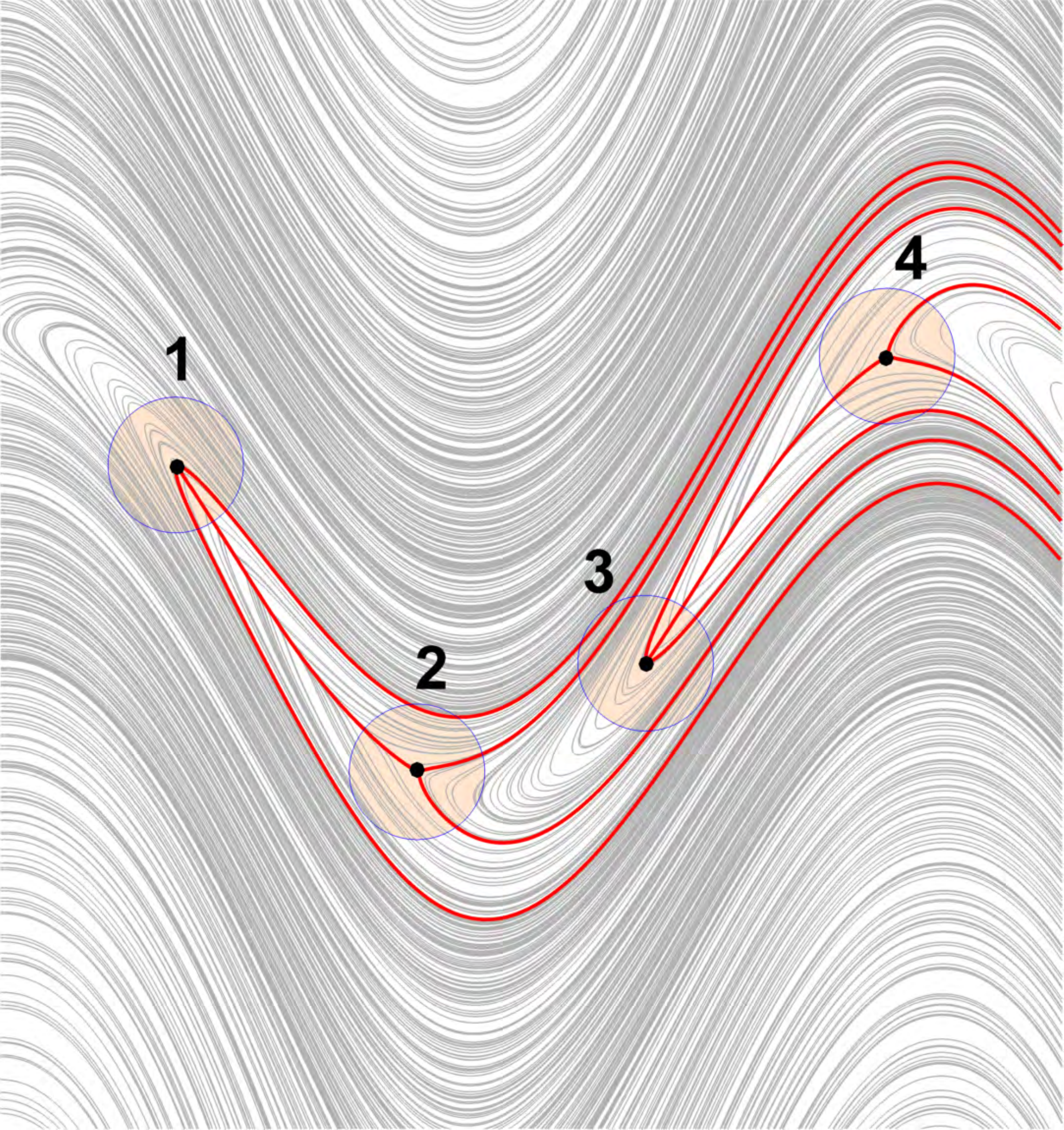}}
\quad{} 
\subfloat[\label{fig:singularities_v2}]{\includegraphics[width=0.3\textwidth]{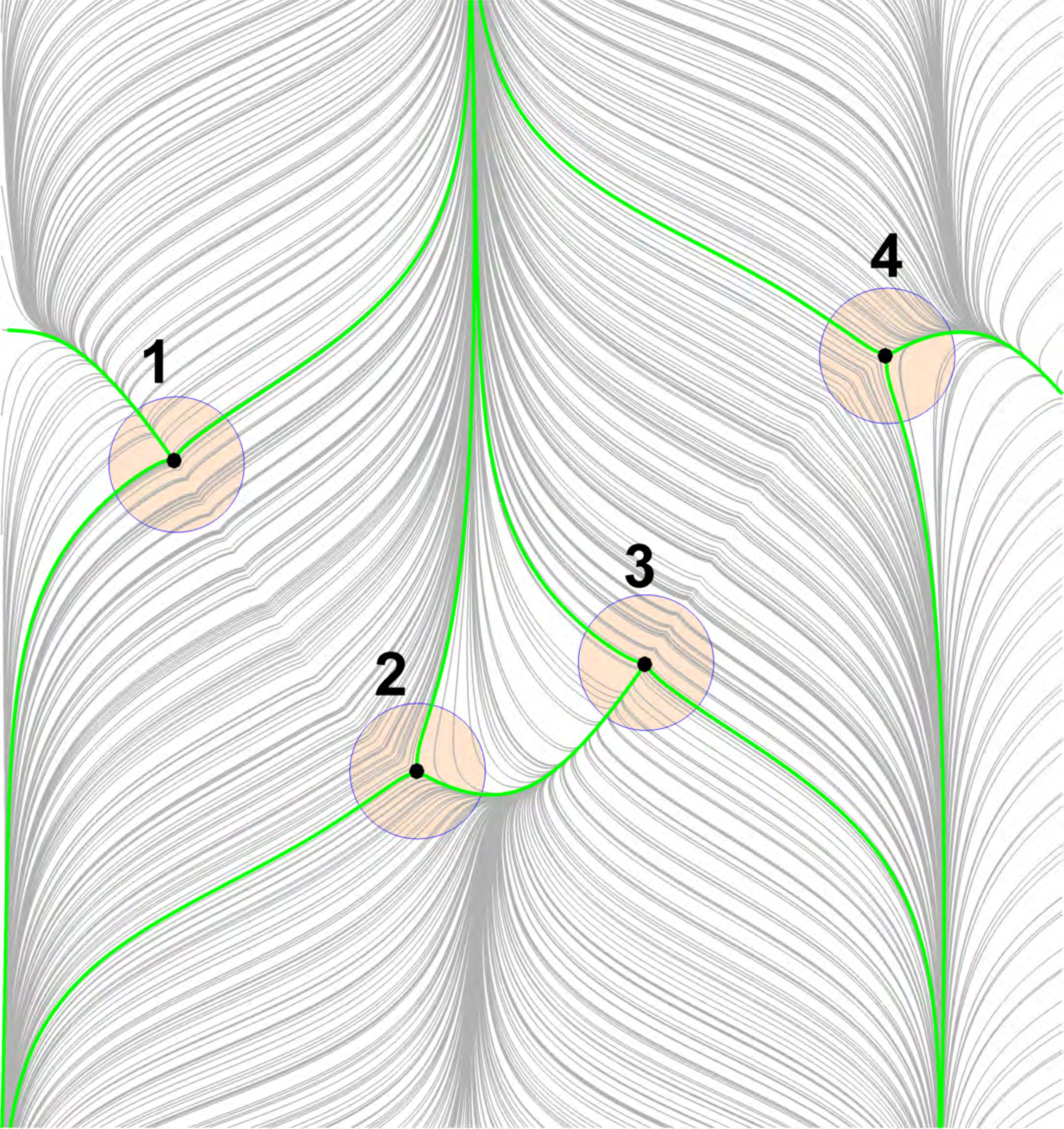}}
\quad{} 
\subfloat[\label{fig:singularities_jet_core}]{\includegraphics[width=0.3\textwidth]{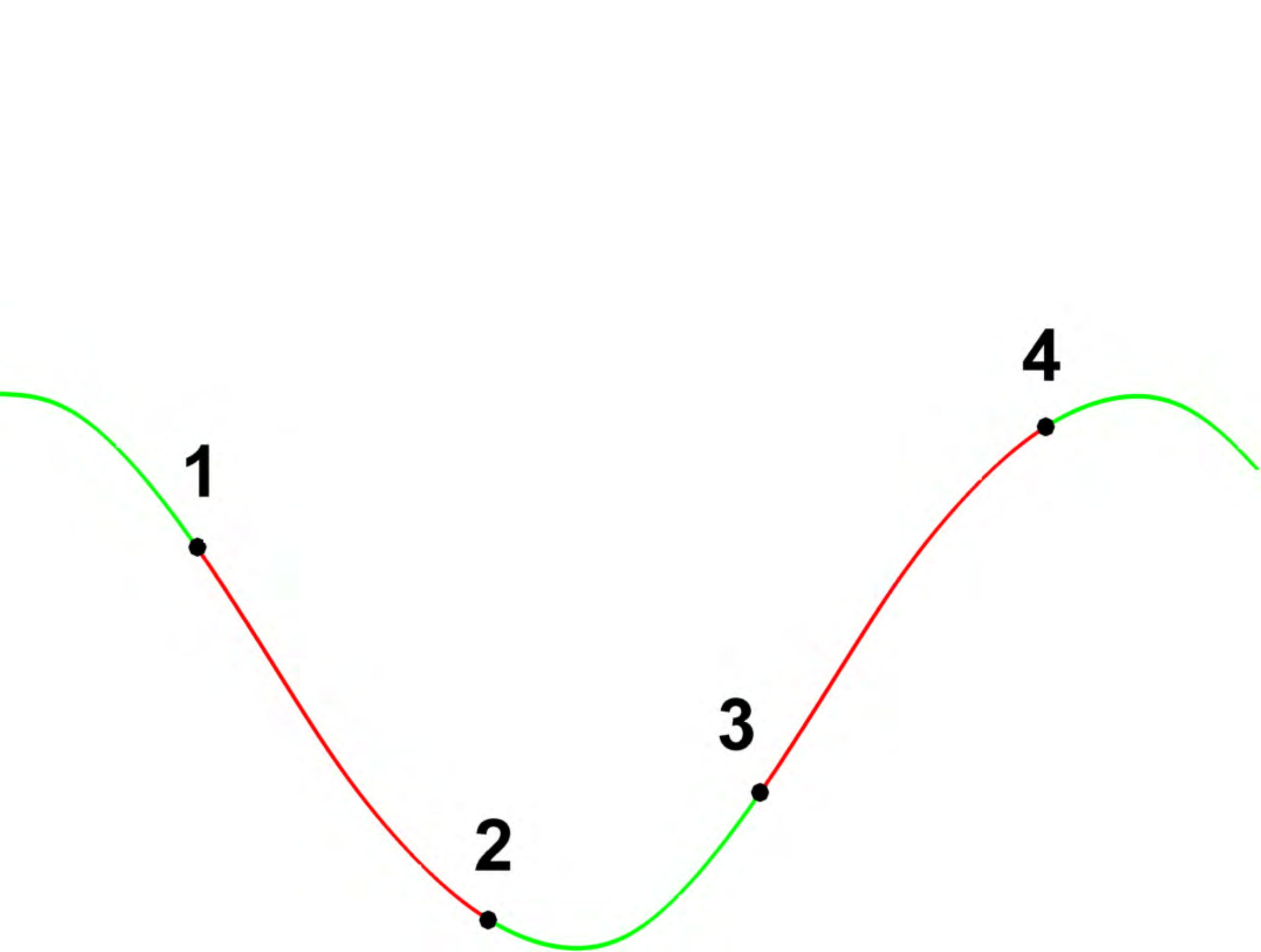}}

\caption{The construction of the initial position of a parabolic LCS for
the standard non-twist map (cf. \cite{faraz13}). (a) Topology of
shrink lines around wedges (singularities 1 and 3) and trisectors
(singularities 2 and 4). Shrink-line separatrices are shown in red.
(b) Topology of stretch lines around singularities. Stretch-line separatrices
are shown in green. (c) Parabolic LCSs formed by a smooth shrink-stretch
separatrix chain that is a weak minimizer of the neutrality \eqref{eq:neutrality}. }

\label{fig:parabolic} 
\end{figure}

Figure \ref{fig:parabolic} illustrates the construction of the time
$t_{0}$ position of a parabolic LCS. The position of such an LCS
at an arbitrary time $t$ can be found by advecting its $t_{0}$ position
under the flow map $F_{t_{0}}^{t}$. Details on the numerical procedures
involved can be found in \cite{faraz13}. 

The resilience of parabolic LCSs (serving as solutions to \eqref{eq:shearlessvari} even under variations to their endpoints) is in agreement with several numerical studies pointing out the robustness and easy observability of Lagrangian jet cores in unsteady zonal flows \cite{Rypina07a,Rypina07b,Beron_Vera08b}. For a general discussion of importance of jet cores and their impact on global weather patterns, see \cite{Francis12}.

Finally, time $t_0$ positions of {\em hyperbolic LCSs} are defined as Cauchy-Green tensorlines starting from local extrema of the strain-eigenvalue fields $\lambda_i(x_0)$ \cite{faraz13,Haller14}. Hyperbolic LCSs, therefore, are also solutions of the stationarity principle \eqref{eq:shearlessvari} for shearless LCSs, but only with respect to variations leaving their endpoints fixed. This constraint on boundary conditions implies a lower degree of observability for hyperbolic LCSs relative to parabolic LCSs. Furthermore, in the short-time, shear-dominated context of the Jupiter video studied here, we have only found weak normal repulsion and attraction along material lines. We therefore omit a  discussion of hyperbolic LCSs in this study.

\subsection{Elliptic LCSs\label{sub:Elliptic-LCSs}}

By Fig. \ref{fig:curve_deformation}, the averaged Lagrangian repulsion
experienced by a closed material curve $\gamma$ over the time interval
$\left[t_{0},t\right]$ is given by 
\begin{equation}
Q(\gamma)=\oint\displaylimits_{\gamma}q(r(s),r'(s))ds=\oint\displaylimits_{\gamma}\frac{1}{\sqrt{\left\langle r^{\prime}(s),\left[C_{t_{0}}^{t}(r(s))\right]^{-1}r^{\prime}(s)\right\rangle \left\langle r^{\prime}(s),r^{\prime}(s)\right\rangle }}ds.
\end{equation}

To generalize the concept of a Kolmogorov--Arnold--Moser-type (KAM-type)
transport barrier \cite{Arnold89} from time-periodic to finite-time
aperiodic flows, we seek closed material curves $\gamma$ along which
the averaged repulsion $Q$ shows no leading-order variation. Satisfying
\begin{equation}
\delta Q(\gamma)=0,\label{eq:strainlessvari}
\end{equation}
such a closed curve $\gamma$ has a thin annular neighborhood in which
no material filamentation occurs over the time interval $[t_{0},t]$,
just as in neighborhoods of KAM curves in the time-periodic case.
As a result, the interior of $\gamma$ exhibits no advective mixing
with its exterior. The observability of solutions of \eqref{eq:strainlessvari}
is on par with those of \eqref{eq:shearlessvari}. Indeed, they prevail
under the largest possibly set of admissible variations, as long as
those variations are also closed curves. \\

For incompressible flows, stationarity of the averaged normal repulsion
is equivalent to the stationarity of the averaged tangential stretching
defined along the material line $\gamma$. As shown in \cite{black_hole},
the latter problem is solved by closed null-geodesics of the Lorentzian
metric family 
\[
g_{\lambda}(u,v)=\left\langle u,E_{\lambda}v\right\rangle ,\quad\lambda>0,
\]
with the generalized Green--Lagrange strain tensor $E_{\lambda}(x_{0})$
defined as 
\[
E_{\lambda}(x_{0})=\frac{1}{2}\left[C_{t_{0}}^{t}(x_{0})-\lambda^{2}I\right].
\]

The metric $g_{\lambda}$ is Lorentzian (i.e., indefinite) on the
set $U_{\lambda}=\left\{ x_{0}\in U\,:\,\lambda_{1}(x_{0})<\lambda^{2}<\lambda_{2}(x_{0})\right\} .$
In this set, all closed null-geodesics of $g_{\lambda}$ are trajectories
of one of the two families of ODEs 
\begin{equation}
r^{\prime}(s)=\eta_{\lambda}^{\pm}(r(s)),\qquad\lambda\in\mathbb{R}^{+},\label{eq:lambdaODE}
\end{equation}
where 
\begin{align*}
\eta_{\lambda}^{\pm}(r) & =\sqrt{\frac{\lambda_{2}(r)-\lambda^{2}}{\lambda_{2}(r)-\lambda_{1}(r)}}\xi_{1}(r)\pm\sqrt{\frac{\lambda^{2}-\lambda_{1}(r)}{\lambda_{2}(r)-\lambda_{1}(r)}}\xi_{2}(r).\\
\end{align*}
Again, for reasons of observability, we focus on structurally stable
closed trajectories (i.e., limit cycles) of \eqref{eq:lambdaODE}.

\begin{figure}
\centering
\includegraphics[width=0.6\textwidth]{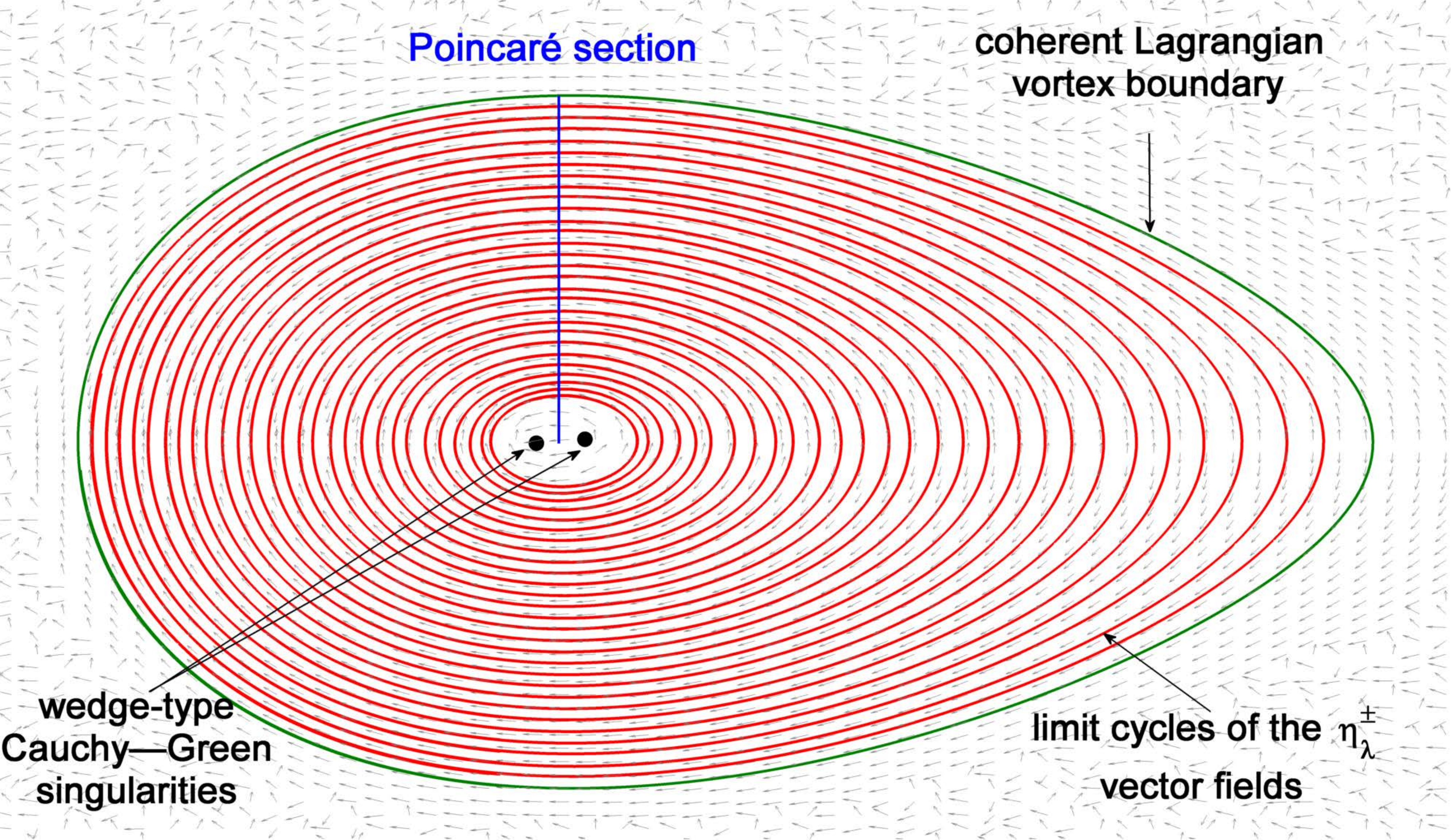}
\caption{The construction of a Lagrangian vortex boundary
for the periodically forced Duffing oscillator \cite{Hadjighasem}.
Elliptic LCSs are identified as limit cycles of the $\eta_{\lambda}^{\pm}$
direction fields via a Poincare section. The outermost elliptic LCS
is highlighted with green as the Lagrangian vortex boundary.}
\label{fig:elliptic} 
\end{figure}

Time $t_{0}$ positions of \emph{elliptic LCSs} are, therefore, limit
cycles of the line fields \eqref{eq:lambdaODE}. Such limit cycles
turn out to encircle at least two wedge-type singularities of the
Cauchy--Green strain tensor field (Fig. \ref{fig:elliptic}). This
fact enables the automated numerical detection of elliptic LCS even
in complex flow fields \cite{Karrasch14b}.

The position of an elliptic LCS at an arbitrary time $t$ can be found
by advecting its $t_{0}$ position under the flow map $F_{t_{0}}^{t}$.
Any limit cycle $\gamma$ of \eqref{eq:lambdaODE} turns out to be
uniformly stretching under such advection \cite{black_hole}. This
means that the arclength of any subset of $\gamma$ increases exactly
by the factor $\lambda$ under the flow map $F_{t_{0}}^{t}$. Limit
cycles of \eqref{eq:lambdaODE} only tend to exist for $\lambda\approx1$,
guaranteeing a high degree of material coherence for a \emph{coherent
Lagrangian vortex boundary}, defined in \cite{black_hole} as the
outermost member of a nested family of limit cycles of \eqref{eq:lambdaODE}.

Figure \ref{fig:elliptic} illustrates the construction of the
time $t_{0}$ slice of an elliptic LCSs and coherent Lagrangian vortex
boundaries. Details on the numerical procedures involved can be found
in \cite{black_hole}.

\section{Unsteady transport barriers in the atmosphere of Jupiter}
\subsection{Prior work and present objectives}
\label{sub:Prior_work}
Notable earlier attempts to identify transport barriers in Jupiter's
atmosphere from images started in \cite{Dowling88}, with velocities
inferred from a manually assisted image-correlation analysis of 10
pairs of photos. The resulting velocities were then all viewed as
part of an underlying steady wind field. This steady vector field
produced spiraling streamlines near the GRS without a clear indication
of material barriers.

Later work \cite{Choi07} used three high-resolution snapshots from
the 2000 Galileo Mission to construct a steady velocity field from
automated cloud-tracking via image correlation velocimetry. Again,
trajectories spiraling into the GRS emerged from this time-averaged
analysis, indicating no particular closed material barrier around
the GRS. More recently, a potential-vorticity-conserving flow (steady
in the frame co-moving with the GRS) was constructed in \cite{Shetty10}
as a best fit to cloud motion inferred from different space missions.
By construction, this averaged approach renders all streamlines closed
near a vortical feature (such as the GRS). The approach, however,
does not address the question of actual material transport via the
unsteady winds of Jupiter.

Extracting an unsteady velocity field and analyzing its finite-time transport barriers has not been attempted in prior publications. One reason for this is a clear focus of the planetary science community on long-term evolution in Jupiter's climate. Comparing velocity snapshots and averages taken from different missions and different years, rather than studying a video footage from a single mission, is clearly more appropriate for a study of climate evolution. The unavoidable time-dependence of velocities extracted from temporally close video frames has, in fact, been viewed as undesirable uncertainty to an envisioned steady mean velocity field (see, e.g., Asay-Davis et al. \cite{ACCIV}). Another reason for the lack of unsteady transport barrier studies for Jupiter has been the unavailability of precise mathematical tools (such as those surveyed here in Section \ref{sec:LCS}) for LCS extraction from finite-time, aperiodic velocity data.

\subsection{Video footage}
\label{section:Video_footage}
The raw footage acquired by the Cassini Orbiter comprises 14 cylindrical
maps of Jupiter, covering the 10 days ranging from October 31 to November
9 in the year 2000. We use an enhanced
version of this video, which NASA created by interpolation
and by addition of information from previous Jupiter missions \cite{NASA_link1,NASA_link2}.
The pixels of the enhanced footage extend to 360 degrees of longitude
and 180 degrees of latitude, with a resolution of $3601\times1801$.
The time step between the frames is 1.1 hours. 

\begin{figure}
\centering
\subfloat[]{\includegraphics[width=0.45\textwidth]{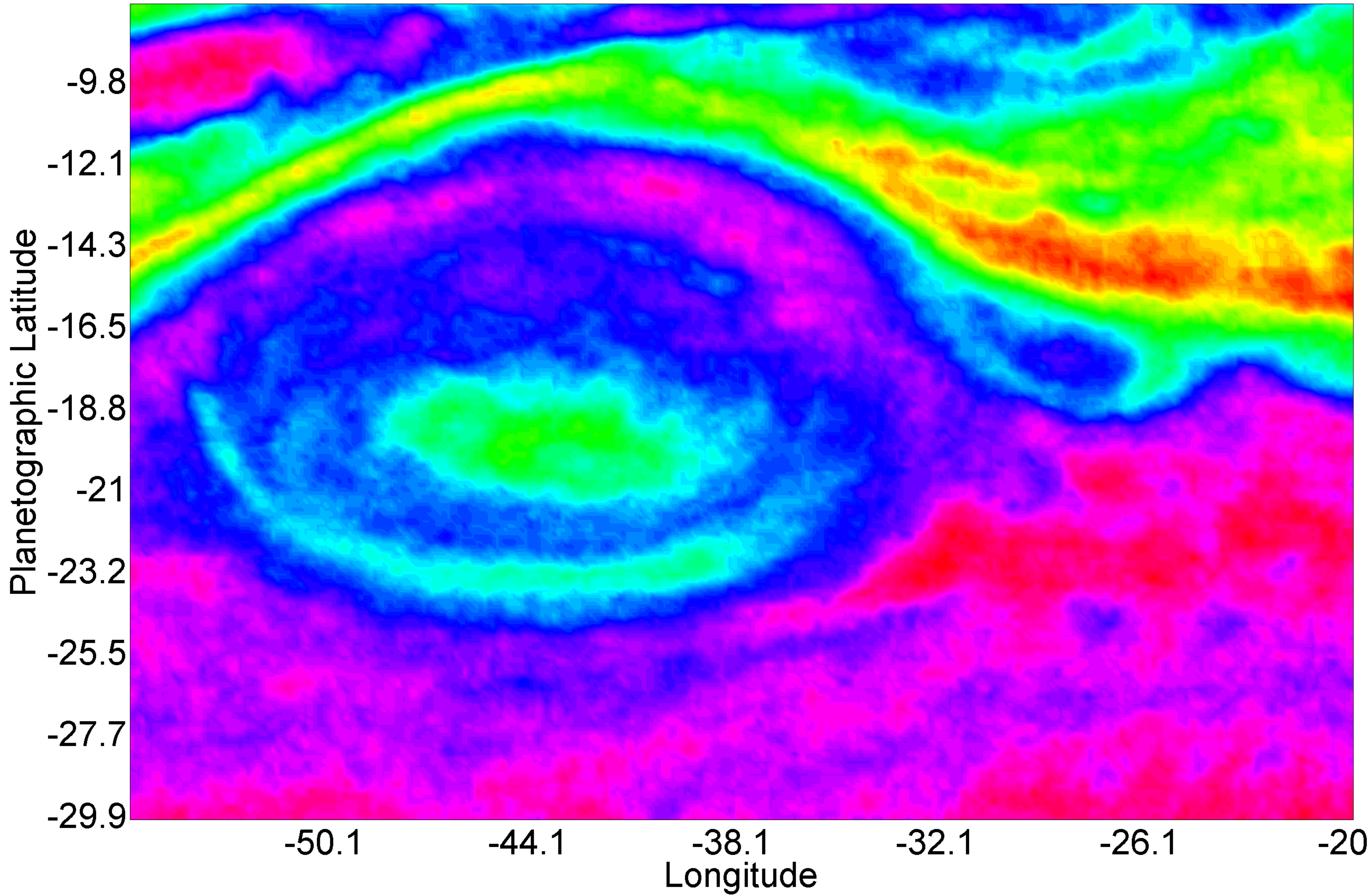}}
\quad
\subfloat[]{\includegraphics[width=0.45\textwidth]{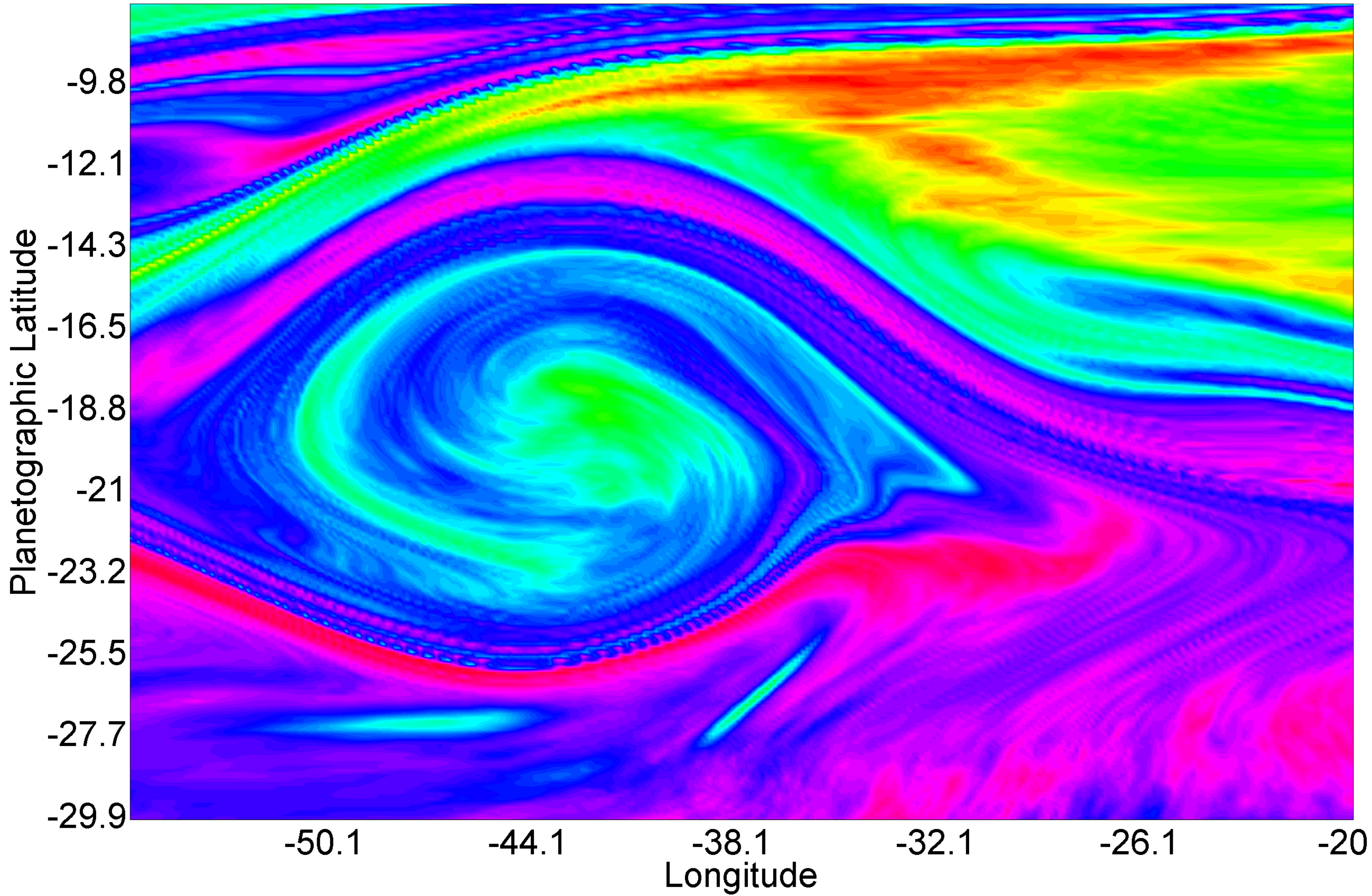}}
\caption{(a) Final frame of the raw footage. The image is shown in Hue-Saturation-Value (HSV) colormap. (b) The first frame of the raw footage is advected to the final time using the extracted velocity field.}
\label{fig:advection}
\end{figure}

To verify the feasibility of the enhanced cloud
movie for Lagrangian advection studies, we used the extracted velocity
field (to be described below) to advect the first image of the raw
footage up to the final time of the same footage. We show representative portions of these two images in Fig. \ref{fig:advection} for comparison. We computed an offset error as the $l^{2}$-distance
between the pixels of the advected initial raw image and the raw image at the final
time, normalized by the $l^{2}$-norm of the final raw image. In this
fashion, we obtained an offset error of $5.6\%$, which is in the
order otherwise expected from numerical noise, processing errors, and diffusive cloud mixing.

\subsection{Optical velocity field reconstruction}

Image-correlation analysis applied to available
cloud videos provides a two-dimensional representation of Jupiter's
winds. Early wind-field reconstruction studies required a human operator
to identify the same cloud feature in subsequent frames \cite{Mitchell}.
The seminal paper by Limaye \cite{Limaye} introduced the first automated
image-correlation method using one-dimensional correlations along
latitudinal circles. Later improvements involved two-dimensional extensions
and coarse-to-fine iteration schemes \cite{Bruck}. Recent approaches
add an advection equation to the procedure \cite{ACCIV}, or apply
the idea of an optical flow \cite{Liu}. Similar methods have also
been applied to image sequences of other planets such as Saturn \cite{Sanchez-Lavega11},
Uranus and Neptune \cite{Kaspi}, and the Earth \cite{Leese}.

Here we use the Advection Corrected Correlation Image Velocimetry
(ACCIV) algorithm developed in \cite{ACCIV} to extract a time-resolved
atmospheric velocity field from the enhanced Cassini footage described in section \ref{section:Video_footage}. ACCIV
uses the idea of the two-pass Image Correlation Velocimetry (CIV) developed
by \cite{Tokumaru,Fincham97,Fincham00} for experimental fluid velocity measurement.
In \cite{ACCIV}, the ACCIV algorithm was employed to reconstruct
steady velocity fields from image pairs provided by the Hubble Space
Telescope, as well as by the Galileo and Cassini space missions. To
our knowledge, however, the ACCIV algorithm has not been employed
before to reconstruct and analyze a time-resolved, unsteady velocity
field from a full video footage.

ACCIV is an iterative technique for reconstructing
a velocity field that is assumed to advect an observed scalar field
passively. As a first step, ACCIV recursively splits two successive
images $I$ and $J$ into sub-images or correlation boxes. Then, for
a correlation box $C_{i}\subset I$, ACCIV finds a matching
box $C_{j}\subset J$ of the same size. The process
of matching correlation boxes is performed by maximizing the cross-correlation
between intensity patterns. The velocity at the center of $C_{i}$
is then the distance between the centers of $C_{i}$ and $C_{j}$
divided by the time elapsed.

\begin{figure}
\begin{centering}
\texttt{ACCIV first pass} 
\par\end{centering}
\centering 
\subfloat[]{\includegraphics[width=0.43\textwidth]{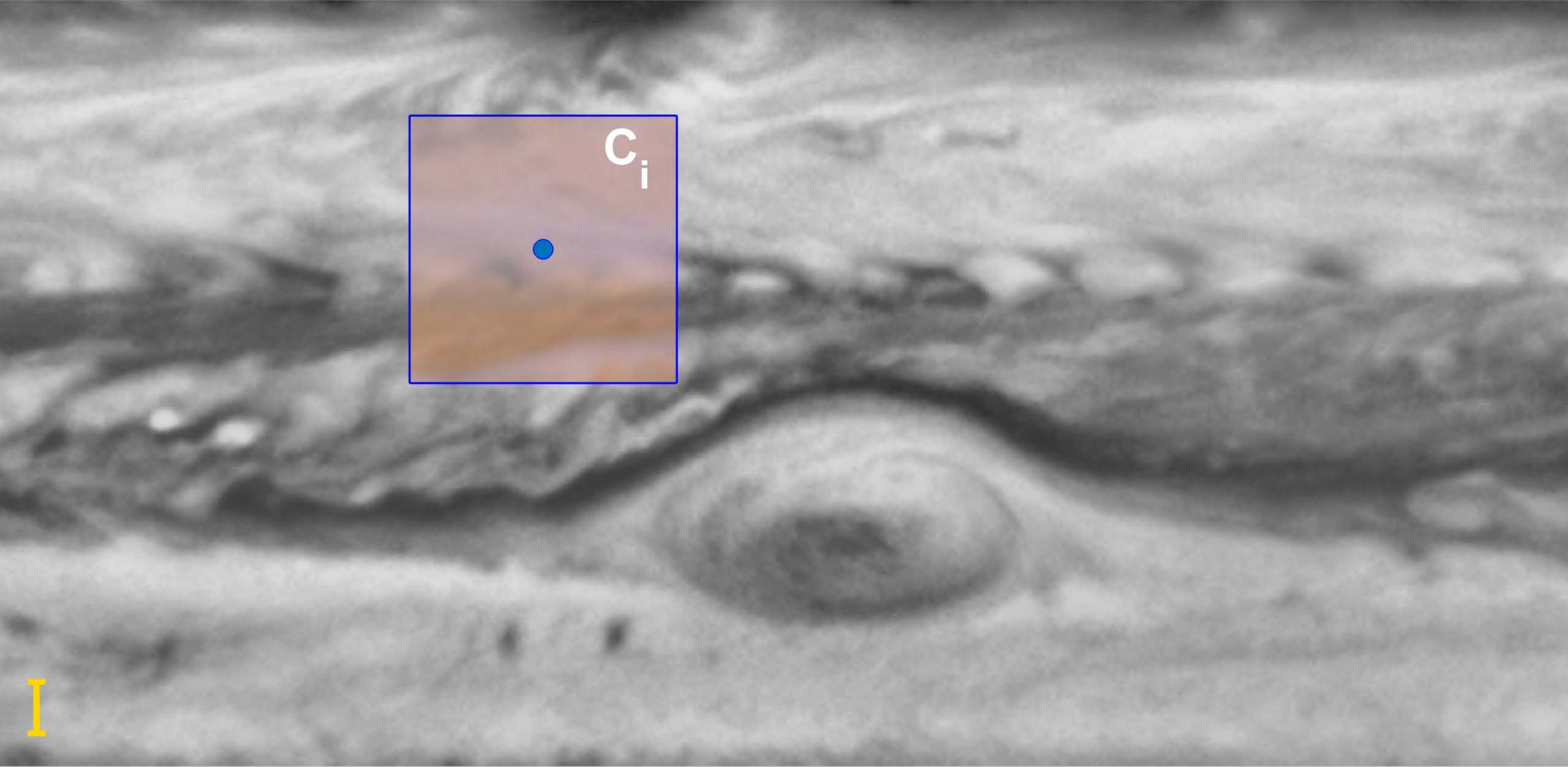}}
\quad{}
\subfloat[]{\includegraphics[width=0.43\textwidth]{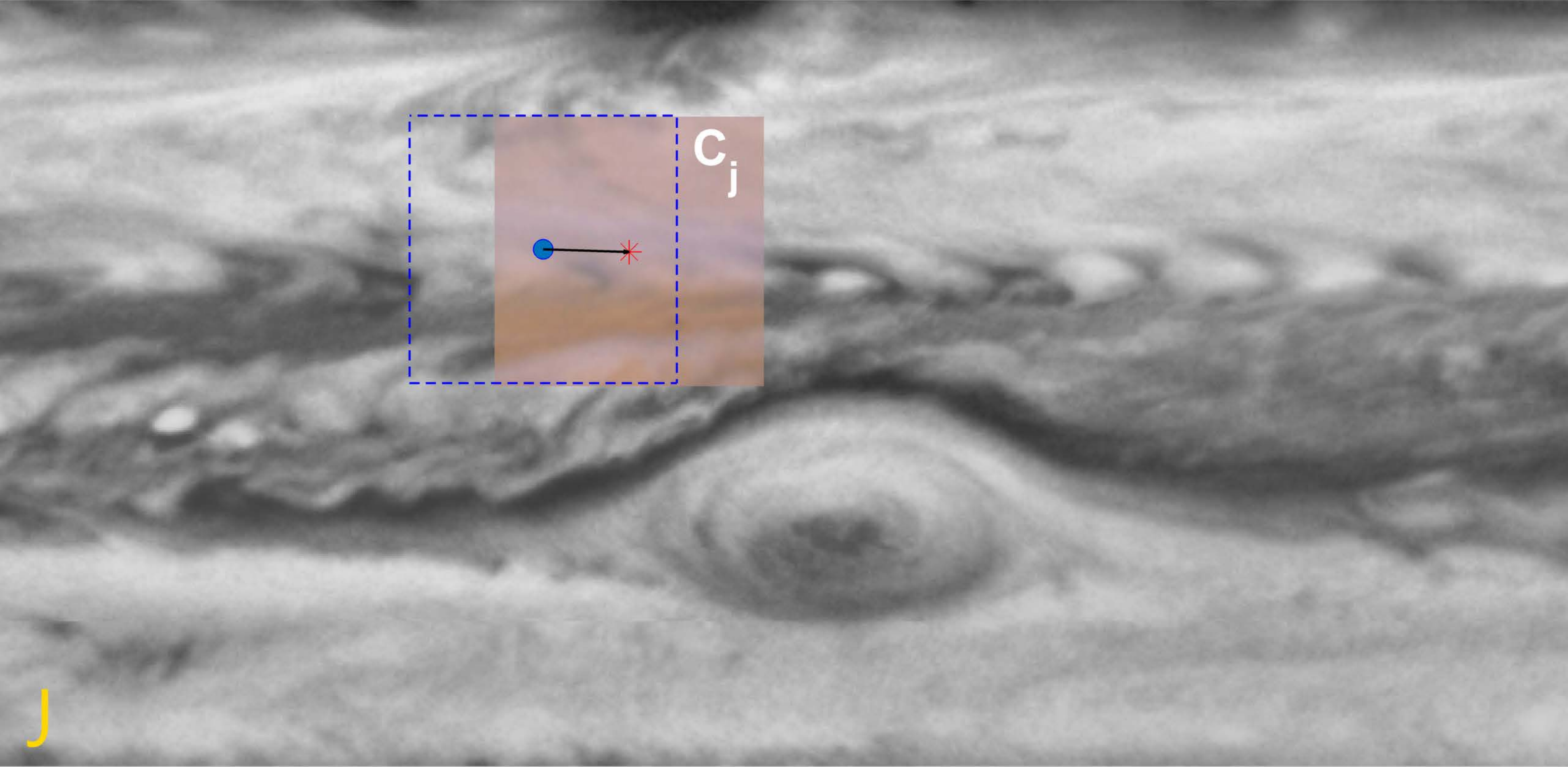}}

\begin{centering}
\texttt{ACCIV second pass} 
\par\end{centering}
\centering 
\subfloat[]{\includegraphics[width=0.43\textwidth]{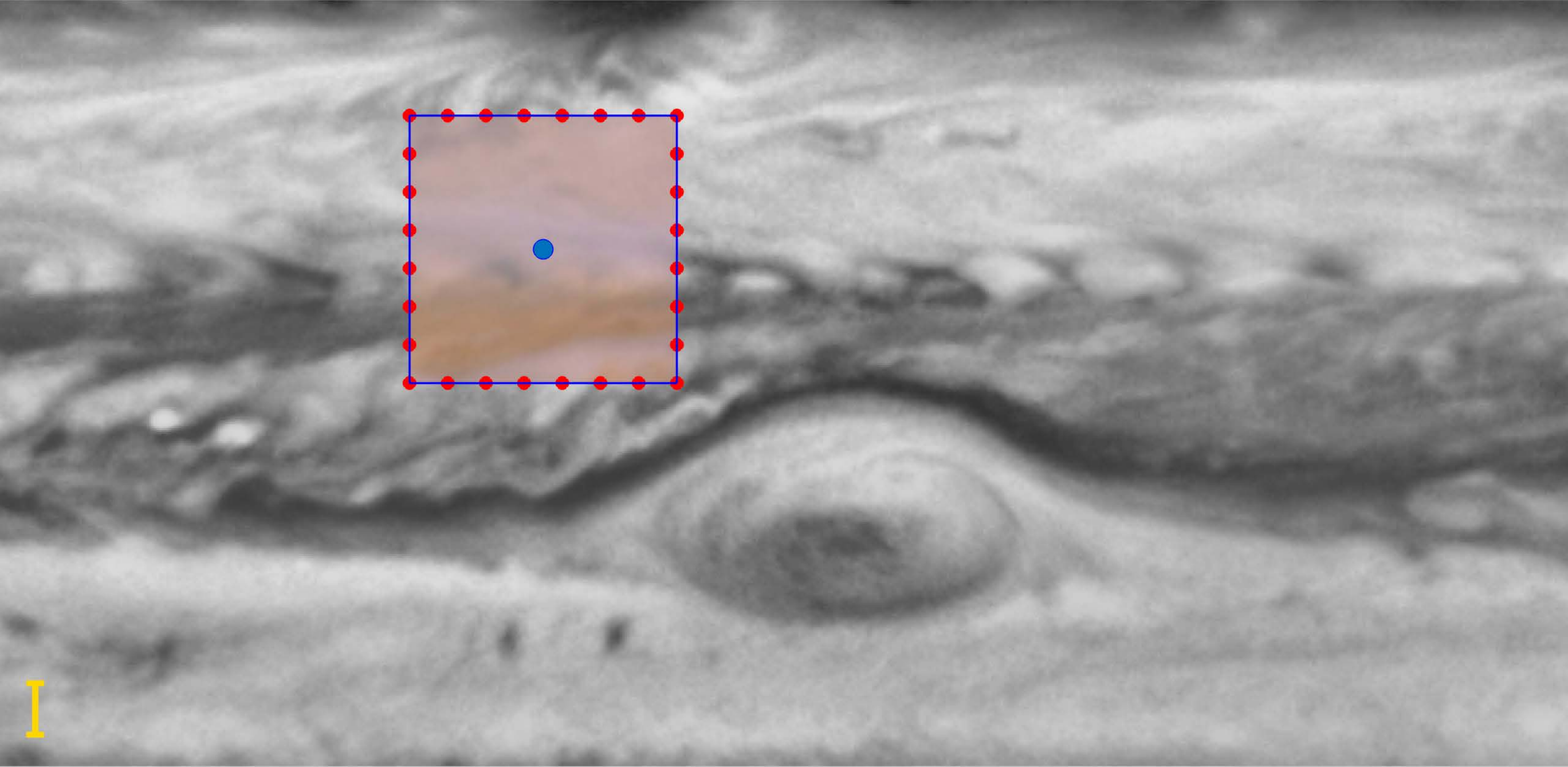}}
\quad{}
\subfloat[]{\includegraphics[width=0.43\textwidth]{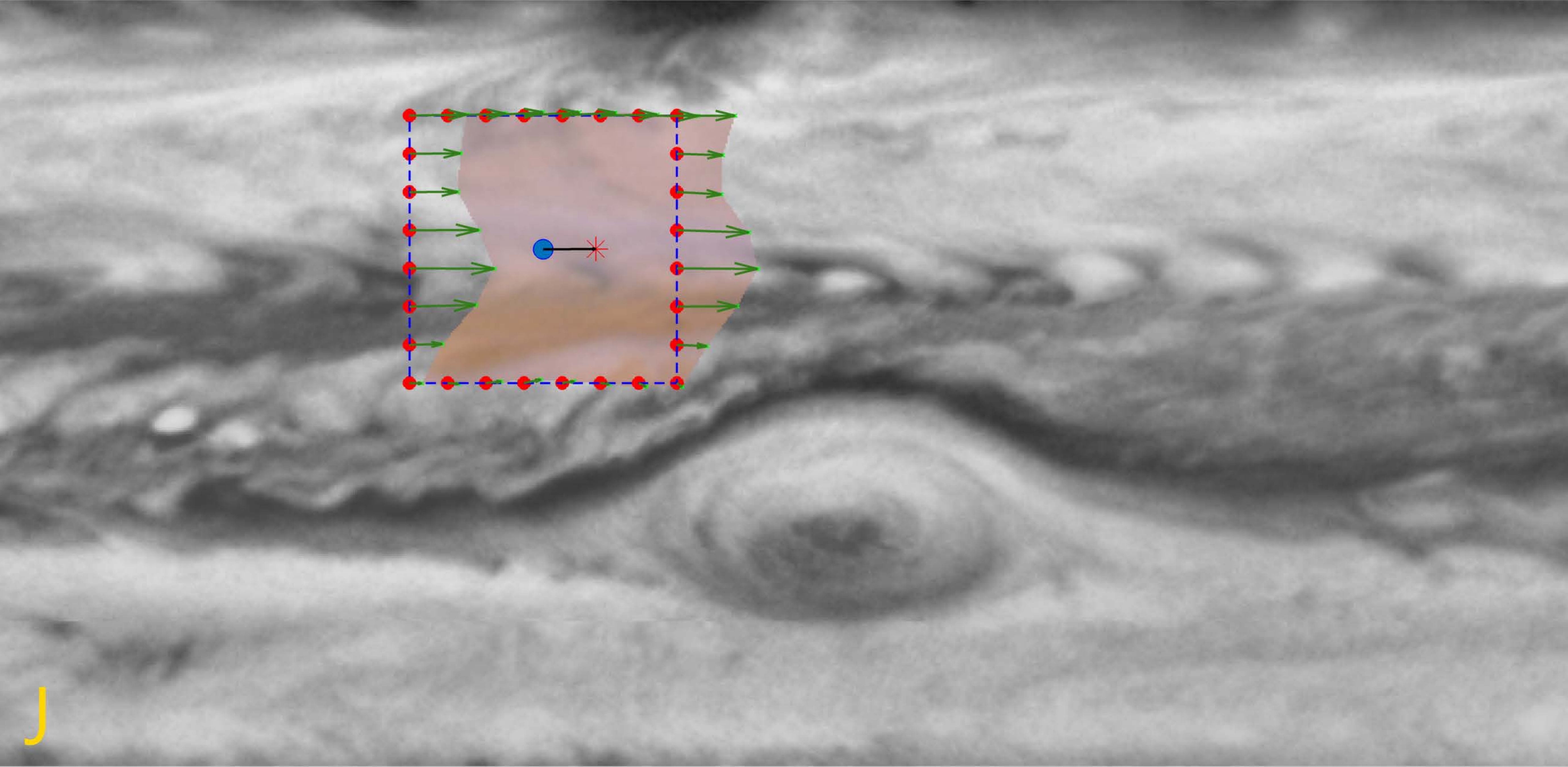}}
\caption{Correlation boxes in the first and second ACCIV
passes. (a) A correlation box of pixels $C_{i}$ in image $I$ is
shown as a solid line. (b) The matching correlation box $C_{j}$ in
the subsequent image $J$ is shown along the silhouette of the correlation
box $C_{i}$. The black arrow connecting the centers of the correlation
boxes shows the estimated displacement vector at the center of $C_{i}$.
The displacement vector is obtained under the assumption that $C_{i}$
moves from image $I$ to $J$ without any distortion. (c) The correlation
box of pixels $C_{i}$ at initial image $I$. (d) The distorted matching
correlation box in the image $J$. The given freedom to the correlation
box $C_{i}$ leads to a better estimation of the displacement vector
(shown in black).}
\label{fig:correlation_box} 
\end{figure}

The algorithm repeats this process for all the correlation
boxes in the image $I$, yielding a crude velocity field approximation
under the assumption that a correlation box moves from image $I$
to the subsequent image $J$ without any distortion. The crude initial
velocity approximation is then used to advect the images to some intermediate
time when no real data are available. The difference between the synthetic
images at the intermediate time is iteratively used to generate correction
vector fields, producing increasingly accurate velocity vectors. In
a second step, ACCIV makes use of the first-step results and looks
for a correlation between a box of pixels in the image $I$ and a
box of pixels transformed in the subsequent image $J$. The possible
transformations are a combination of translation, rotation, shear
and distortion. Figure \ref{fig:correlation_box} illustrates how
considering a deformed correlation box can lead to a better approximation
of the displacement vector at the center of the correlation box $C_{i}$.
Similar to the first step, ACCIV iteratively improves the accuracy
of the velocity vectors by building synthetic images and producing
correction vectors. The next steps consist of further refinement of
the velocity field using smaller correlation boxes.\\

\begin{SCfigure}
\centering 
\includegraphics[width=0.6\textwidth]{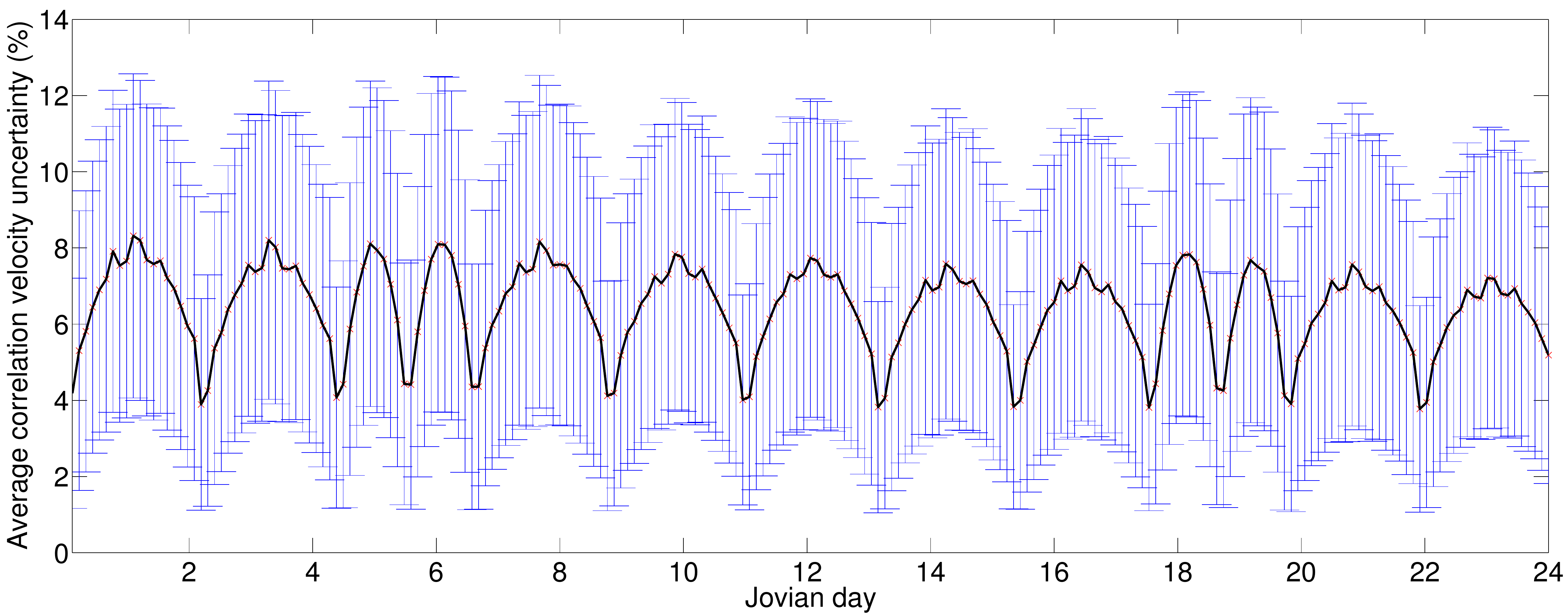}
\caption{The average correlation velocity uncertainty as a percentage of the
spatial mean of the reconstructed velocities is shown in black for
each video frame. The standard deviation of velocity uncertainty data
is shown in blue.}
\label{fig:temporal_uncertainty} 
\end{SCfigure}

\begin{SCfigure}
\centering 
\includegraphics[width=0.6\textwidth]{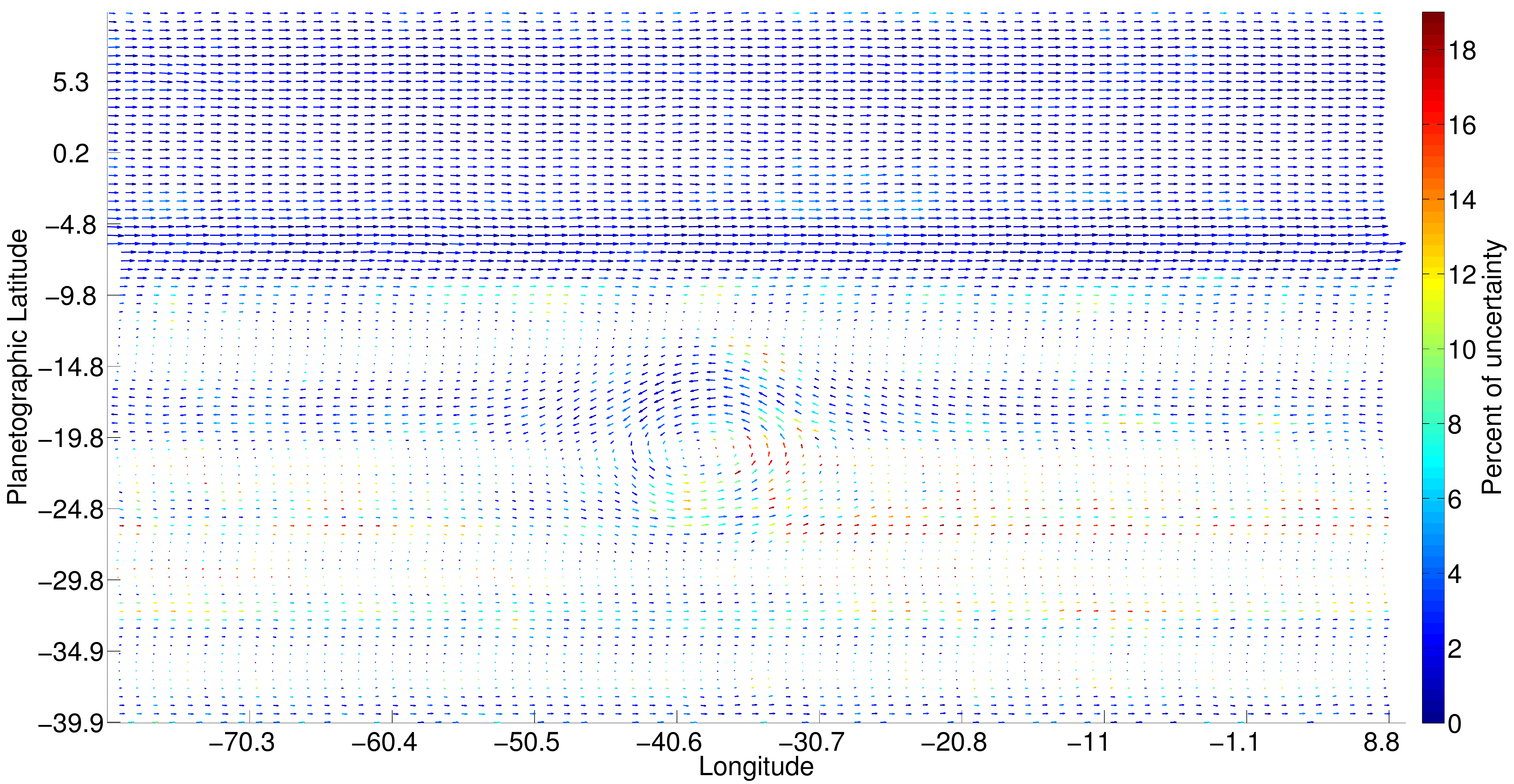}
\caption{The correlation velocity uncertainty is shown as a percent of velocity
magnitude at the location of velocity grid points. The result is shown
for the frame number 105.}
\label{fig:spatial_uncertainty} 
\end{SCfigure}

ACCIV repeats each of these steps iteratively until the \textit{velocity correlation uncertainty} shows no further decrease. First, ACCIV defines the offset between advected and raw images at time $t_{2}$ as local \textit{correlation location uncertainty}. The correlation velocity uncertainty is then the correlation location uncertainty divided by the time elapsed between the images. The velocity fields we have extracted with ACCIV have tens to hundreds of thousands of independent velocity vectors with uncertainties on the order of 2-5 $m/s$.

Figure \ref{fig:temporal_uncertainty} shows the measure of correlation velocity uncertainty as a percentage of the spatial mean velocity for each video frame. The near-periodicity of the uncertainty history arises from the mostly even sampling frequency (roughly two jovian days) of the original raw footage from Cassini. Two time intervals (around days 5-7 and 16-18) with shorter sampling times create an impression of approximate symmetry in the uncertainty distribution with respect to day 12, but this is accidental.

On average, the extracted velocity vectors have about $6.4\%$ uncertainty with respect to the mean of the reconstructed velocities. Figure \ref{fig:spatial_uncertainty} shows the spatial distribution of velocity uncertainty as a percentage of the velocity norm at each grid point. The velocity uncertainty is higher in regions such as the GRS where we have complicated dynamics and presumably cloud mixing.

We recall that we integrate the reconstructed velocity field to find structurally stable structures: limit cycles for elliptic LCSs and robust heteroclinic cycles for parabolic LCSs (cf. Section 3). These structures persist in the (unknown) true velocity field, as long as the imaging and reconstruction errors represent an overall moderate perturbation to the flow map, such as the perturbation we infer from Fig. 5.  This persistence result holds even if the velocity errors are pointwise large at times (cf. \cite{Haller02}). 

Some input parameters for ACCIV, such as the times between frames
and the threshold for removing outliers, are straightforward to specify.
Other parameters, such as correlation box size, search range, and
stride, must be optimized iteratively. Table \ref{table:ACCIV_para}
summarizes the input parameters and results for each pass. For more
information on setting the input parameters of ACCIV, we refer the
reader to the project webpage \cite{ACCIV_patameters}.

\begin{table}[t]
\centering
\caption{ACCIV parameters used to produce the time-resolved velocity field.
The box size, search range, and stride are in units of pixels. The
box size is the size of the correlation box for the relevant CIV pass.
The search range is the range of correlation box displacements used
in each dimension. The stride is the number of pixels by which the
correlation box is shifted between each measurement. It controls the
output resolution of the velocity vector field. The number of image
pairs is the total number of pairing of the set of images. The Number
of Smooth Fit Control Point Neighbors controls the smoothness of the
velocity field on the grid.}

\includegraphics[width=1\textwidth]{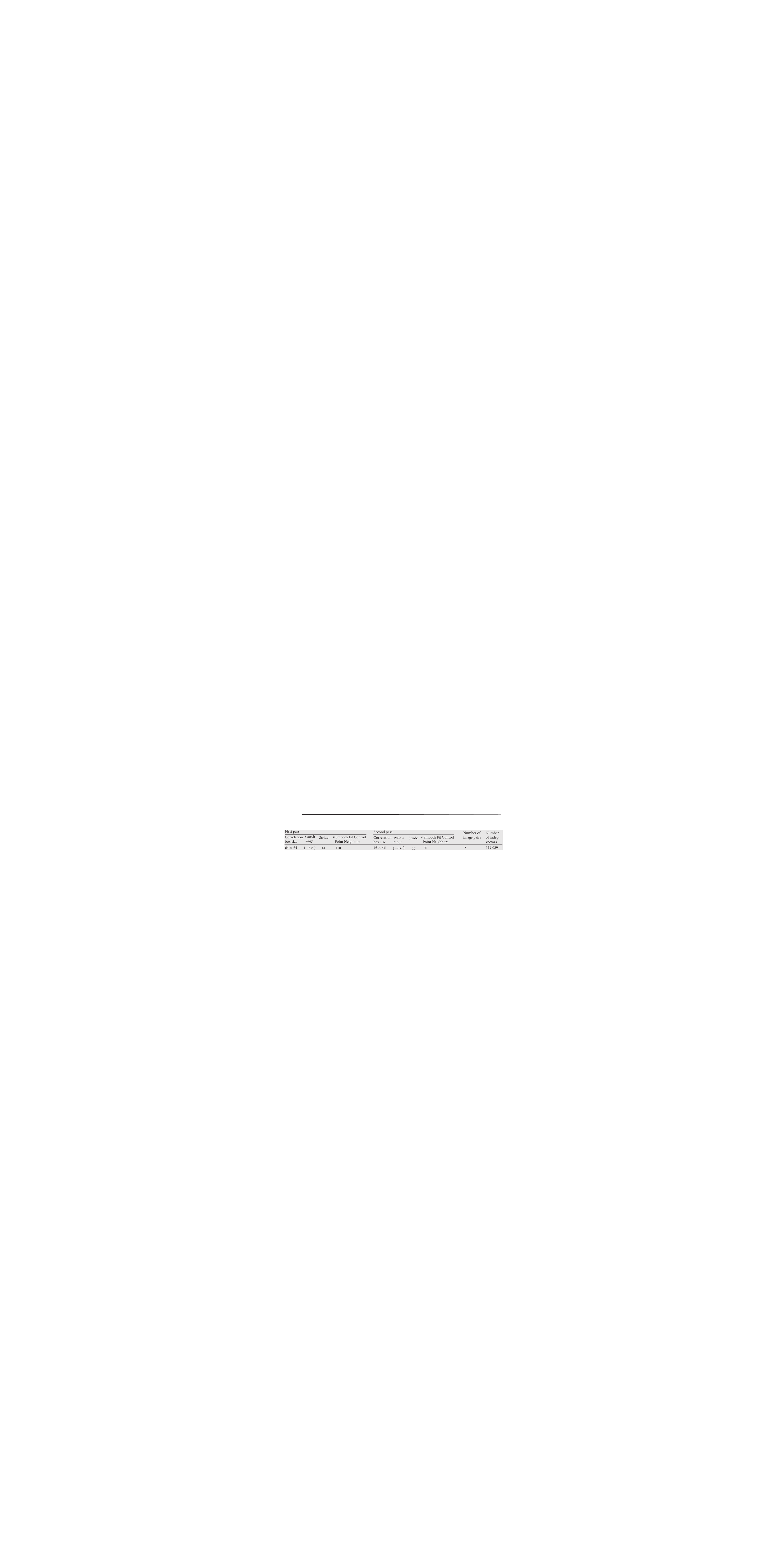} \label{table:ACCIV_para} 
\end{table}

\begin{figure}
\centering 
\includegraphics[width=0.8\textwidth]{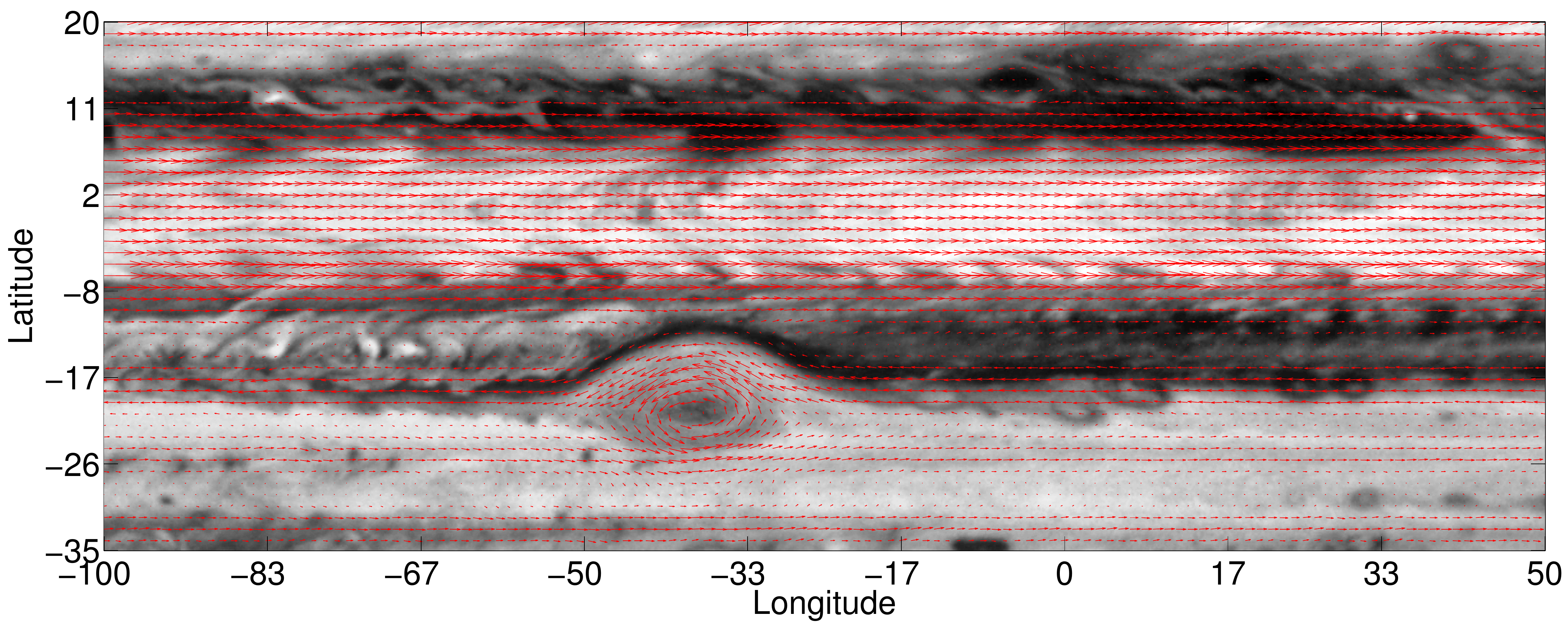}
\caption{A representative snapshot (frame number 105) of the unsteady velocity
field using the ACCIV algorithm. Only one tenth of the total number
of velocity vectors are shown for the sake of visual clarity.}
\label{fig:vector_field} 
\end{figure}

Figure \ref{fig:vector_field} shows a representative snapshot of
the reconstructed unsteady velocity field, which
is available over the domain ranging from $-180\,^{\circ}{\rm W}$
to $180^{\circ}{\rm E}$ in longitudes and from $-35^{\circ}{\rm S}$
to $20^{\circ}{\rm N}$ in latitudes, with a grid resolution of $7202\times1102$.
This supersedes the resolution of earlier velocity fields reconstructed
for Jupiter from manual cloud-tracking approaches \cite{Mitchell,Dowling88,Simon12,Legarreta,Vasavada}.\\

In principle, an alternative to the ACCIV method used here would be Digital Particle Image Velocimetry (DPIV), which reproduces fluid velocities from highly resolved tracks of luminescent particles in well-illuminated laboratory flows \cite{Heitz10}. Under such conditions, high-speed imaging can reliably detect small particle displacements amidst minimal illumination changes, a basic requirement for the optical flow methods underlying DPIV. In observations of planetary atmospheres, however, these ideal imaging conditions are generally not met. For instance, the sunlight scattered from Jupiter's cloud cover into the camera of a spacecraft tends to vary considerably between two subsequent images.

As a consequence, optical flow methods have not gained popularity in velocity field reconstruction from planetary observations. A rare exception is Ref. \cite{Liu}, which works with a single pair of well-lit, high-resolution images from NASA's Galileo mission, separated by one hour. From this data, \cite{Liu} extracts a single high-quality, steady velocity snapshot. Unfortunately, the data available from the Galileo mission is insufficient for the extraction of a reasonably long unsteady velocity field via the approach of \cite{Liu}.

In comparison to the two high-resolution images used in \cite{Liu}, the more modestly resolved but temporally extended Cassini data set used here offers a clear advantage for unsteady LCS detection. As is now well-established for satellite altimetry maps of the ocean, even without capturing smaller (sub-mesocale and lower) features of a velocity field, one can accurately capture its mesoscale LCSs, which in turn agree with in situ float observations \cite{Beron_Vera13}.

\subsection{Validation of the reconstructed velocity field}

Available observational records of Jupiter go back to the late 19th
century, indicating that Jupiter's atmosphere is highly stable in
the latitudinal direction. The average zonal velocity profile as a
function of latitudinal degree is, therefore, an important benchmark
in assessing the quality of the reconstructed velocity field.

\begin{figure}
\centering 
\subfloat[]{\includegraphics[width=0.8\textwidth]{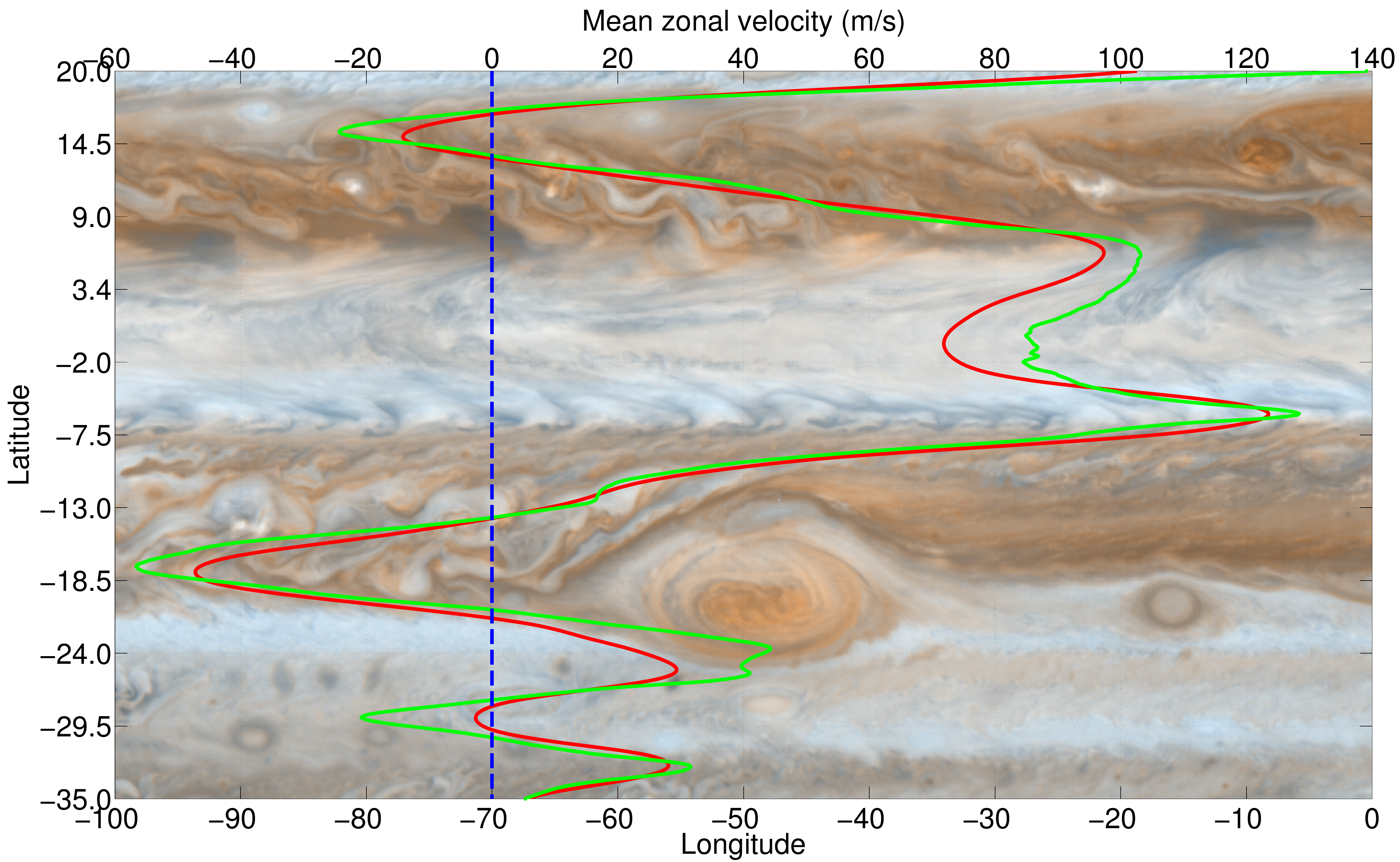}}
\caption{Zonal velocity profile of Jupiter's atmosphere. The red line represents
the average obtained from the reconstructed velocity field; the green
profile is the one reported by S.S. Limaye \cite{Limaye}. The Cassini
map PIA07782 is used as background.}

\label{fig:vel_comparison} 
\end{figure}

In Fig. \ref{fig:vel_comparison}, we compare the temporally averaged zonal velocity profile obtained from ACCIV with the classic profile reported by Limaye \cite{Limaye}. The Limaye profile has been used and confirmed by several other studies on different data sets from different missions (see, e.g., \cite{Ingersoll81,Vasavada,Garcia01}). These studies all support the conclusion that the averaged zonal wind field constructed by Limaye \cite{Limaye} has remained fundamentally unaltered.

Limaye's velocity profile is based on Voyager I and Voyager II images, covering a total of 142 Jovian days in 1979. In contrast, our ACCIV-based velocity profile is based on the Cassini mission, and covers a total of 24 Jovian days in 2000-2001 \cite{NASA_link1}. Overall, the two profiles shown in Fig. \ref{fig:vel_comparison} match fairly closely, showing only appreciable discrepancies near velocity extrema. These discrepancies arise because ACCIV, as other image-correlation methods, systematically underestimates the magnitude of the velocity vectors near peaks of the velocity field \cite{ACCIV}. 

We finally note that we have used images taken at the visible wavelength to extract the velocities of Jupiter's clouds. Asay-Davis et al., \cite{Asay_Davis11} show that velocities extracted from images taken with different visible wavelengths (at different times) produce similar zonal velocities. This observation and other empirical studies support the expectation that Jupiter's cloud velocities can be correctly inferred from images taken at visible wavelengths.

\subsection{Lagrangian advection}

For the parabolic LCS computation described in section \ref{sub:Parabolic-LCSs},
we calculate the Cauchy--Green strain tensor field $C_{t_{0}}^{t}$
defined in \eqref{eq:CG} with $t_{0}=0$ and $t=24$ days, over a
uniform grid $\mathcal{G}_{0}^{1}$ of $5600\times2200$ points. The
spatial domain $U$ ranges from $-145\,^{\circ}{\rm W}$ to $95^{\circ}{\rm E}$
in longitudes and from $-35^{\circ}{\rm S}$ to $20^{\circ}{\rm N}$
in latitudes. For the elliptic LCS computations described in section \ref{sub:Elliptic-LCSs},
a smaller grid $\mathcal{G}_{0}^{2}$ of $900\times600$ points suffices,
because the accurate identification of Cauchy--Green singularities
is not crucial. In all computations, we use a variable-order Adams--Bashforth--Moulton
solver (ODE113 in MATLAB) to solve the differential equations (\ref{eq:dynsys})
and (\ref{eq:lambdaODE}). The absolute and relative tolerances of
the ODE solver are chosen as $10^{-6}$. Off the grid points, we obtain
the $\xi_{j}(x_{0})$ and $\eta_{\lambda}^{\pm}(x_{0})$ line fields
from bilinear interpolation.

\subsection{Unsteady zonal jet cores as parabolic LCSs}

Influential work on model flows has indicated that high potential-vorticity
(PV) gradients occurring along the cores of the eastward jets of Jupiter create effective
barriers to material transport \cite{Juckes87}. More recent work
on perturbed PV-staircase flow models revealed that westward jet cores
also act as material barrier cores, even when the PV has vanishing
gradients along these lines \cite{Beron_Vera08b}. Averaged meridional
velocities support this conclusion \cite{Beron_Vera08b}, but time-resolved
studies using observed winds have not been carried out to ascertain
the existence of actual material transport barriers along eastward
and westward jet cores.

\begin{figure}
\centering 
\includegraphics[width=0.8\textwidth]{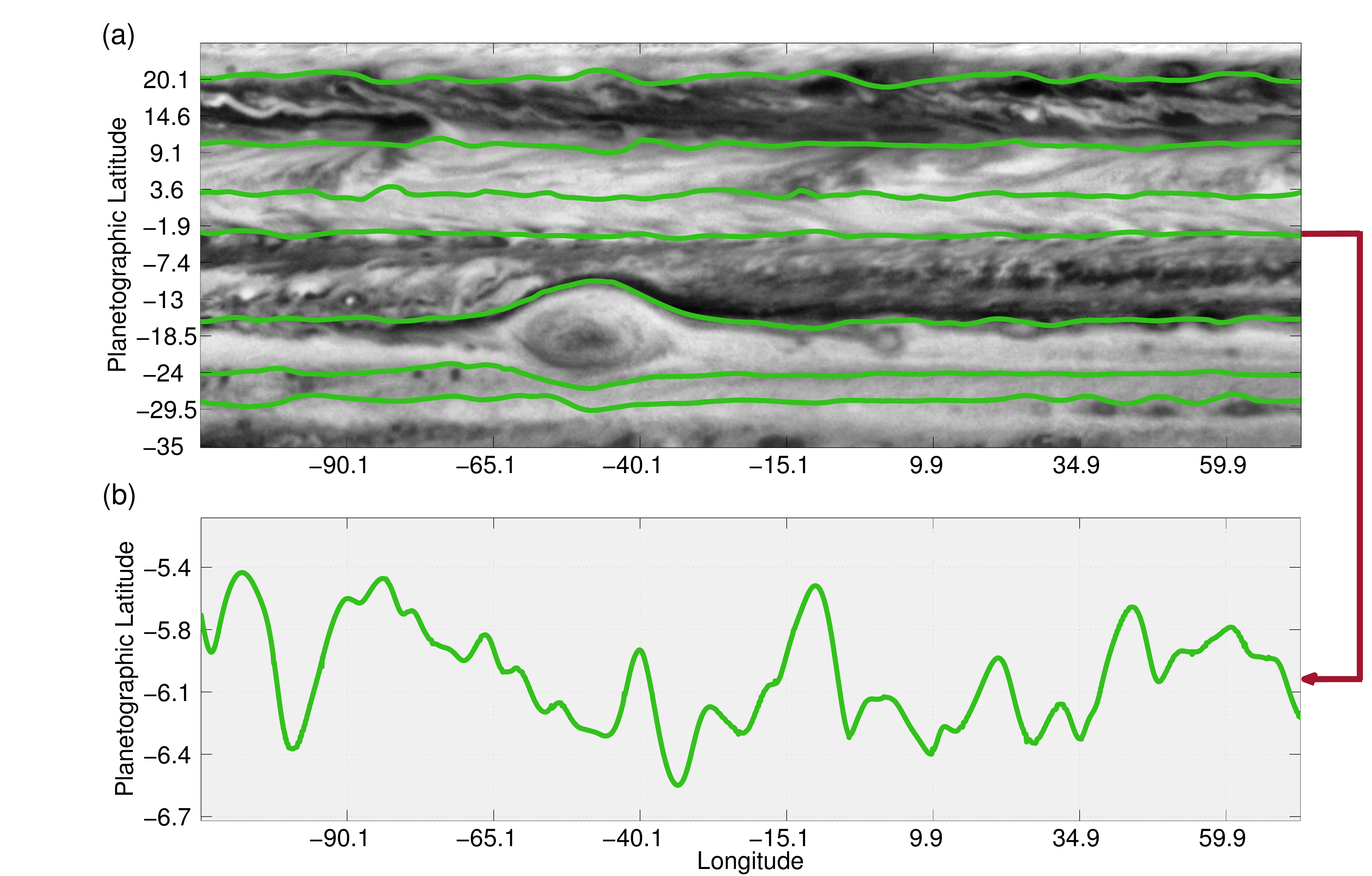}
\caption{(a) Instantaneous positions of parabolic LCSs at the initial time as shearless transport barriers that form the core of jet streams in the atmosphere of Jupiter. (b) The spatial profile of Jupiter's southern equatorial jet. The average meandering width is about $0.33\,^{\circ}$ latitudinal degree.}

\label{fig:Jet_map} 
\end{figure}

Here we examine the validity of the above model-based conclusions
on the unsteady wind field inferred from the time-resolved Cassini
footage. Applying the theory surveyed in section \ref{sub:Parabolic-LCSs},
we compute the Lagrangian cores of jet streams located between latitudes $-35\,^{\circ}{\rm S}$ and $20\,^{\circ}{\rm N}$
(see Fig. \ref{fig:Jet_map}a). In line with the model-based conclusions
of \cite{Beron_Vera08b}, we find coherent Lagrangian jet cores both
for eastward and westward jets. Unlike the straight lines suggested
by individual snapshots, however, the actual unsteady jet cores exhibit
small-amplitude north-south oscillations, as shown in Fig. \ref{fig:Jet_map}b.

\begin{figure}
\centering 
\includegraphics[width=0.7\textwidth]{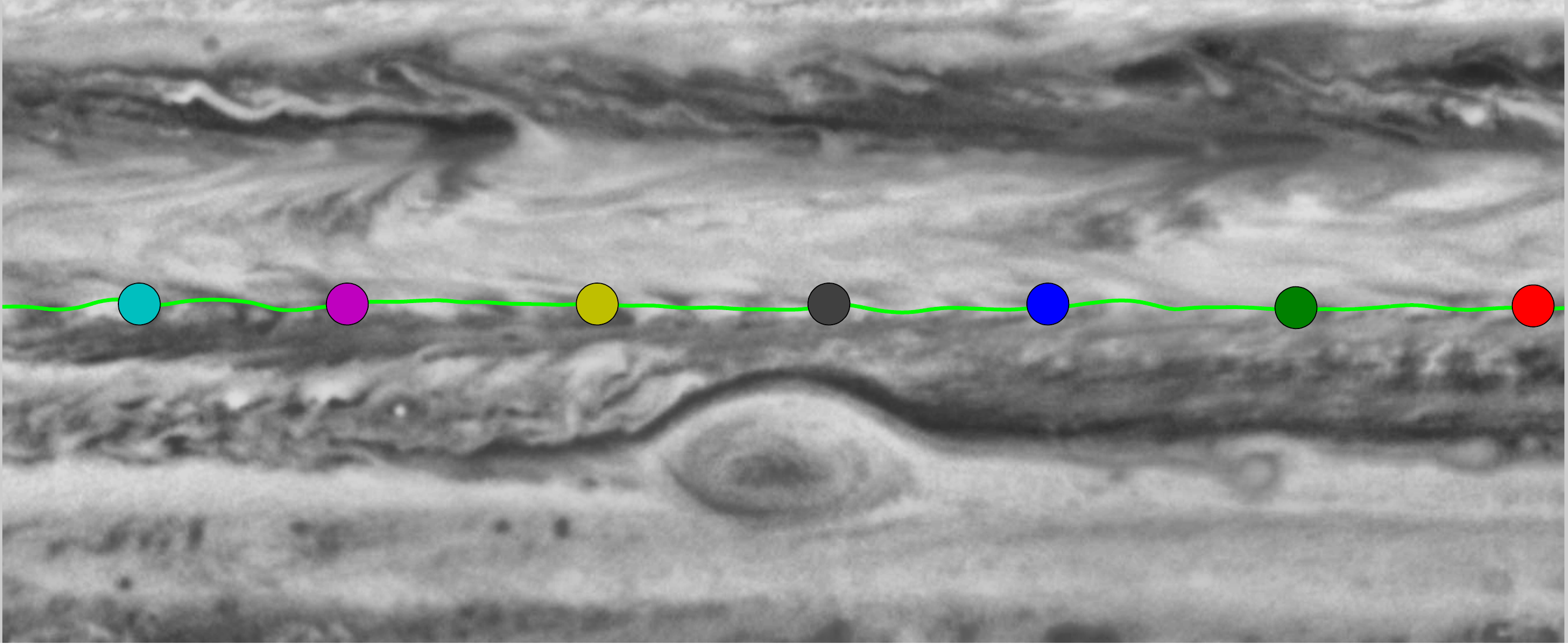}
\quad
\includegraphics[width=0.7\textwidth]{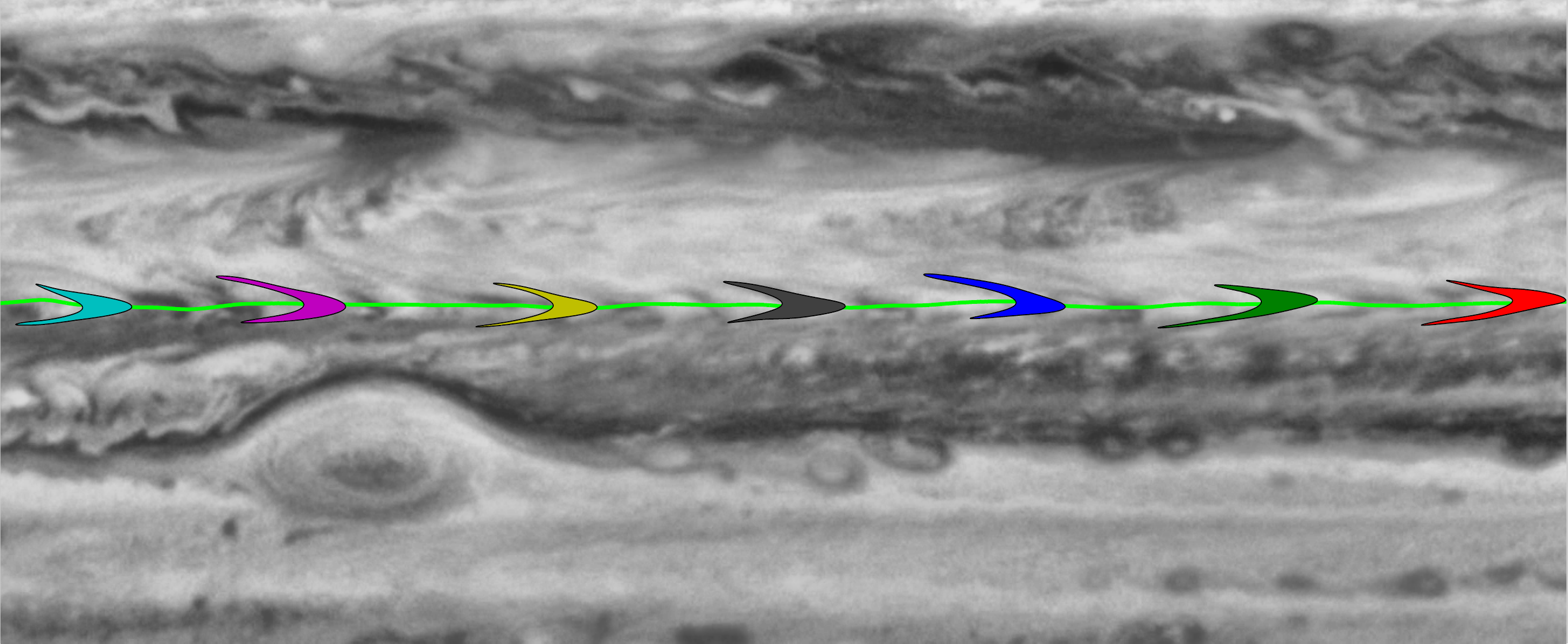}
\caption{Impact of Jupiter's southern equatorial shearless transport barrier on tracer disks over 11 jovian days. The deformed tracer disks resemble the shape of the recently discovered chevrons \cite{Simon12}, i.e., dark v-shaped clouds in the background. The contrast of images is improved for better visualization. The complete advection sequence over 24 Jovian days is illustrated in the online supplemental movie M1.}

\label{fig:chevrons} 
\end{figure}

Figure \ref{fig:chevrons} shows the evolution of initially circular
blobs of tracers, centered on the shearless core of the southern equatorial
jet, after 11 Jovian days. The shape of tracer blobs resembles the
shape of chevrons, serving as Lagrangian footprints of Rossby waves identified recently from an Eulerian perspective on the southern hemisphere of Jupiter \cite{Simon12}.

The advected jet core, as a material line, does not allow mixing between the two wings of any chevron. Specifically, the evolving parabolic LCS does act as a transport barrier, keeping its coherence, and showing no wave-breaking or fingering-type deformation. This coherence effectively blocks the advective excursion of material between the upper and lower half of the material jet.

\subsection{The Great Red Spot as a generalized KAM region}

Observational evidence suggests the existence of coherent rings around
all jovian vortices, including the GRS. The accepted explanation is
that these rings are signs of vertically moving air parcels in the
three-dimensional atmosphere of the planet \cite{Conrath81,Pater10}.
A ring of air parcels is a material transport barrier that is
expected to have a two-dimensional footprint in the horizontal wind
field around the GRS. Such a signature, however, has not been identified
in available advection studies (cf. section \ref{sub:Prior_work})

Here we seek an annular material transport barrier region around the GRS by applying the geodesic LCS theory
described in section \ref{sub:Elliptic-LCSs}. This necessitates the computation of limit cycles for the family of autonomous dynamical systems defined in eq. \eqref{eq:lambdaODE}. We discard limit cycles obtained for the same $\lambda$ value, if they are within two velocity grid steps from each other. This is to focus on robust enough limit cycles that are far enough from undergoing a saddle-node bifurcation.

\begin{figure}[h!]
\centering
\subfloat[\label{fig:closed_orbits_t0}]{\includegraphics[width=0.52\textwidth]{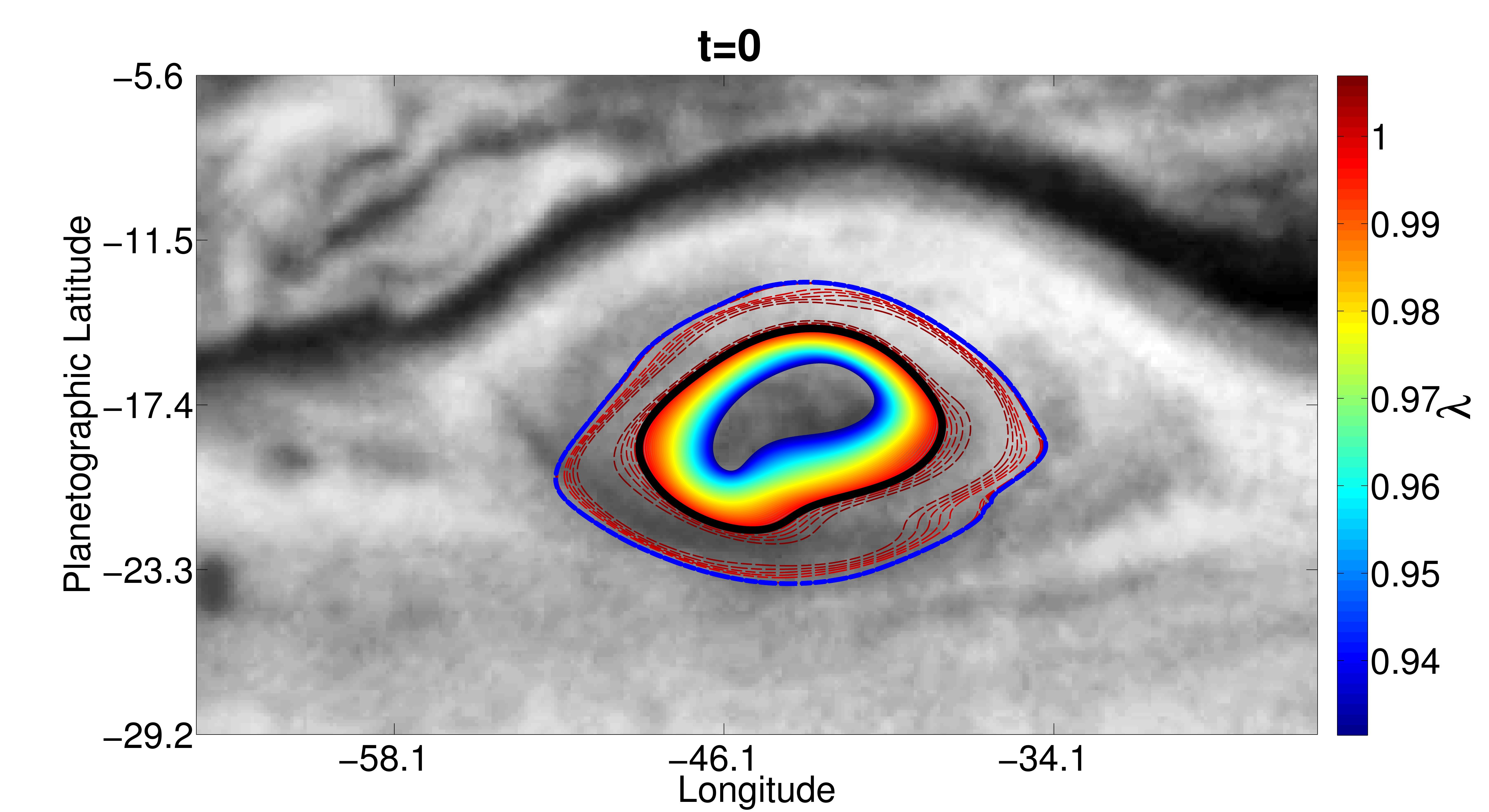}}
\;
\subfloat[\label{fig:GRS_t0}]{\includegraphics[width=0.46\textwidth]{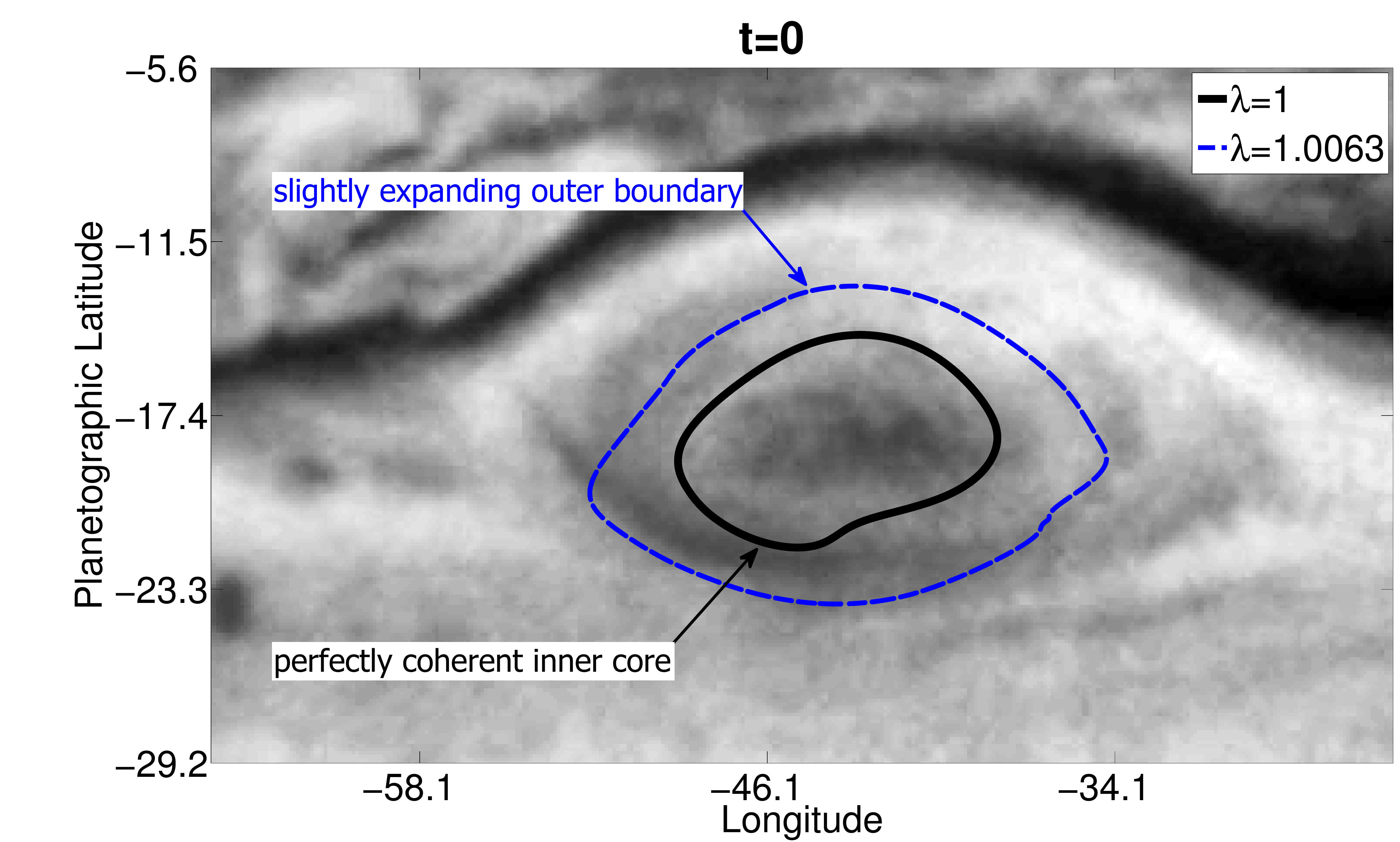}}

\subfloat[\label{fig:closed_orbits_T}]{\includegraphics[width=0.52\textwidth]{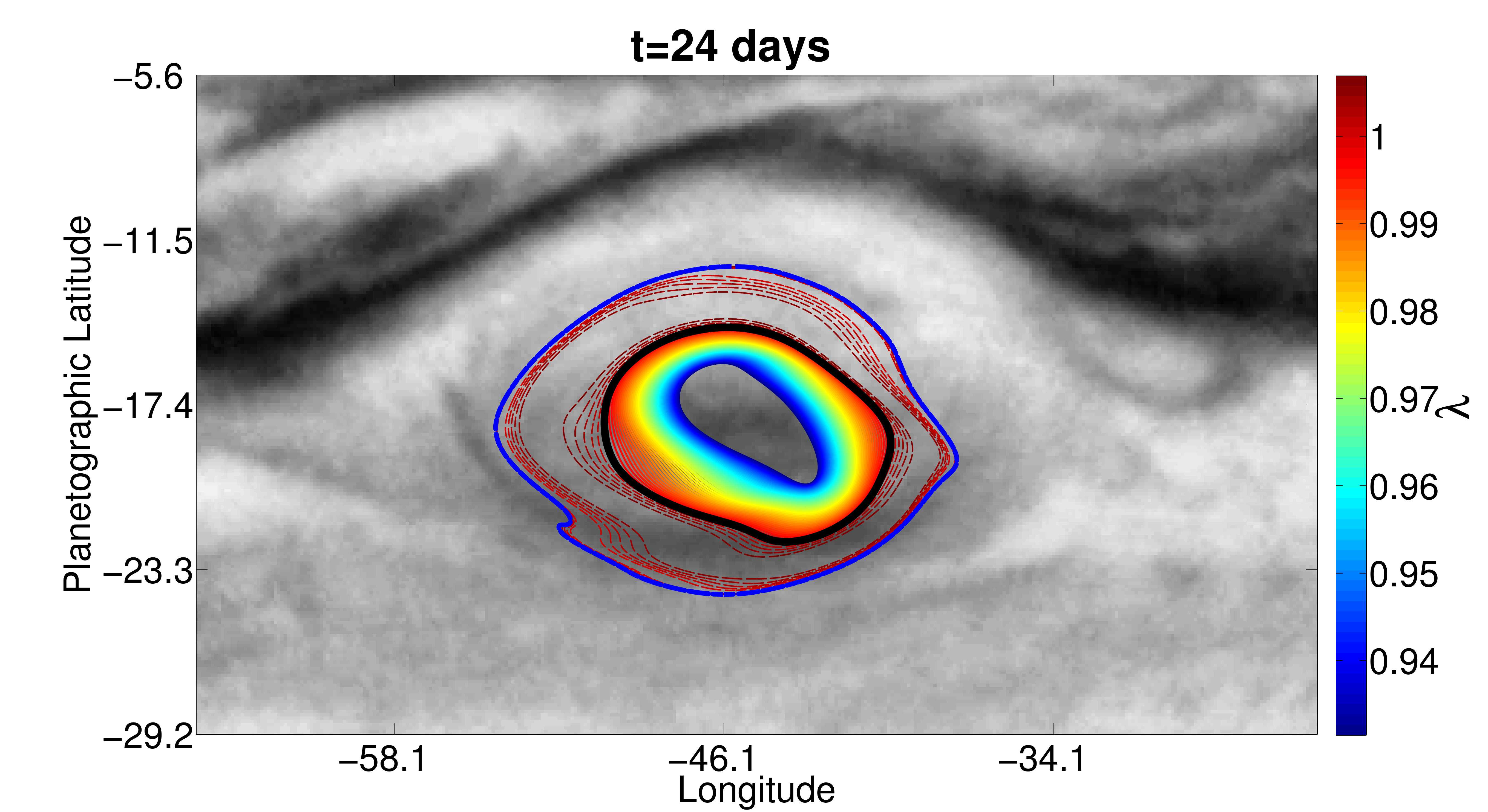}}
\; 
\subfloat[\label{fig:GRS_T}]{\includegraphics[width=0.46\textwidth]{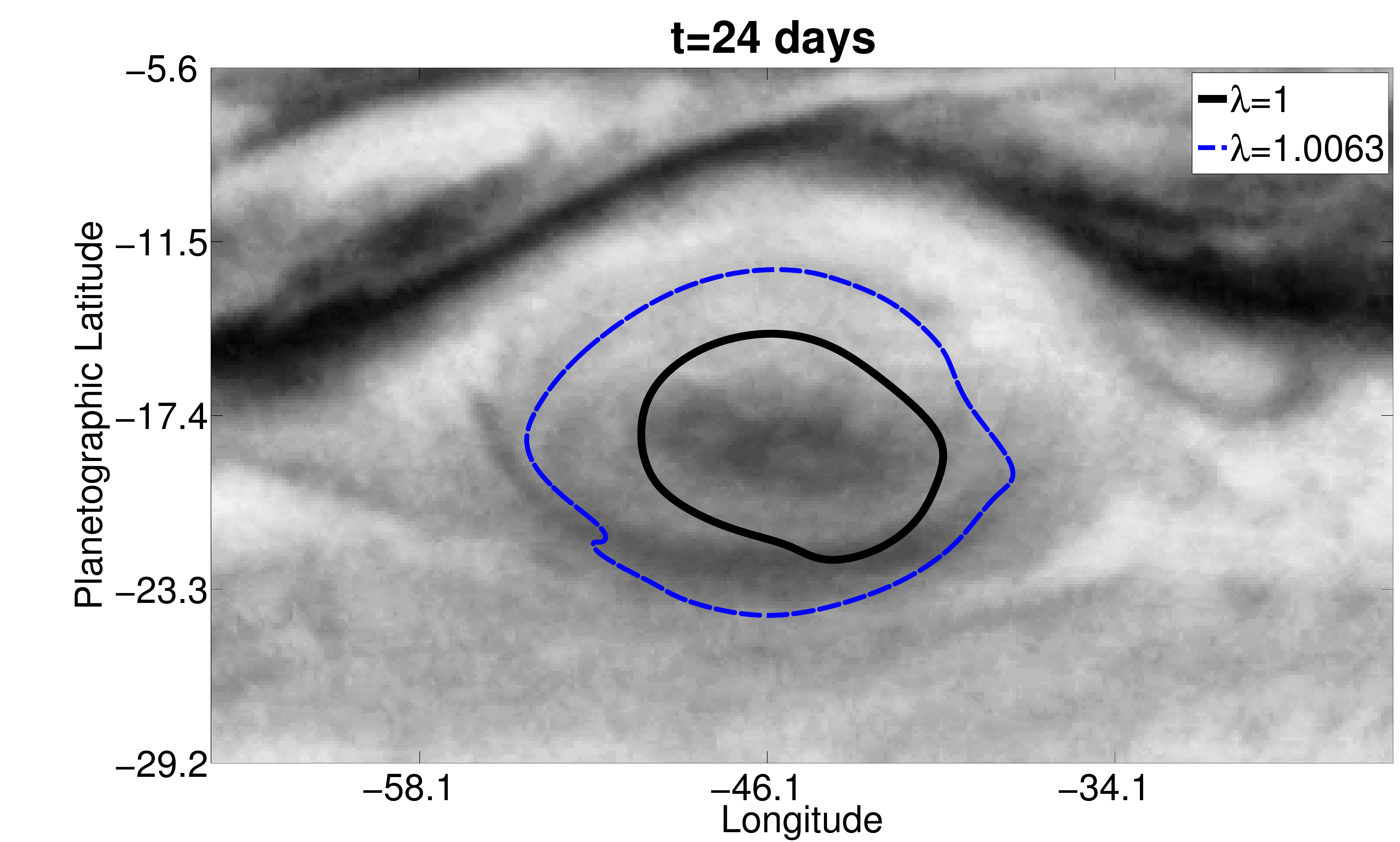}}

\subfloat[\label{fig:length_change}]{\includegraphics[width=0.48\textwidth]{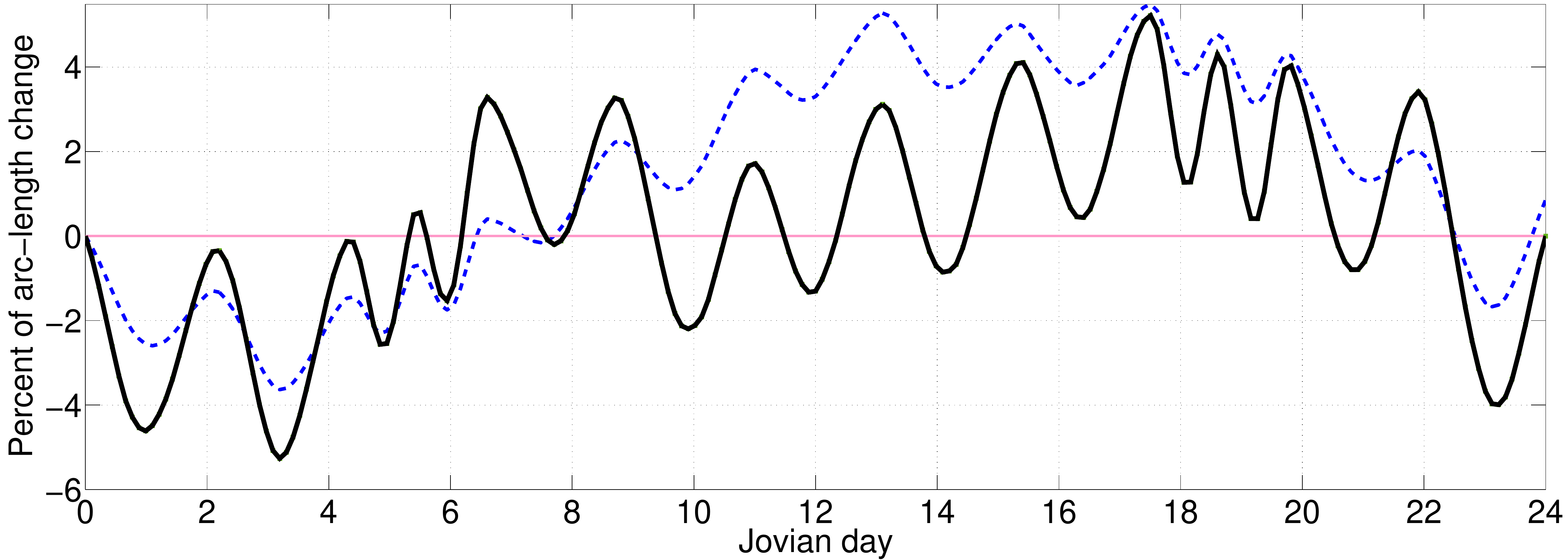}}
\quad{} 
\subfloat[\label{fig:aspect_ratio}]{\includegraphics[width=0.48\textwidth]{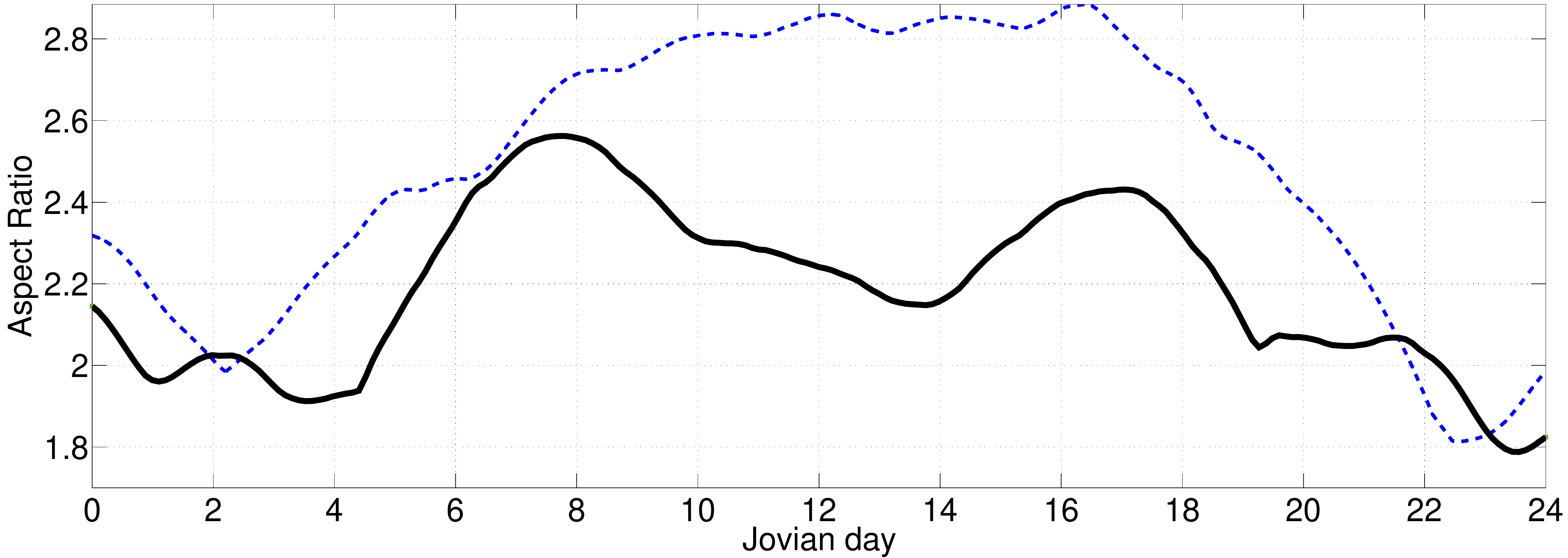}}

\caption{(a) Elliptic LCSs defining the Lagrangian footprint of the Great Red Spot at time $t=0$. The color-bar refers to values of the stretching parameter $\lambda$ arising in the construction of the elliptic LCS family. (b) Elliptic LCS with perfect coherence (black), as well as the Lagrangian vortex boundary (blue) of the Great Red Spot at time $t=0$. The boundary is extracted from velocity data covering 24 Jovian days. The advection sequence is illustrated in the online supplemental movie M2. (c) Elliptic LCSs defining the Lagrangian footprint of the Great Red Spot at time $t=24$. (d) Elliptic LCS with perfect coherence (black), as well as the Lagrangian vortex boundary (blue) of the Great Red Spot at time $t=24$. (e) Relative stretching of the perfectly coherent ($\lambda=1$) inner-core boundary and the slightly expanding outer boundary of the GRS over 24 Jovian days. As predicted by geodesic theory, the arc-length of the outer boundary changes about $0.6\%$ in agreement with the theoretical stretching value $(\lambda=1.0063$) of the extracted outer boundary. (f) plot of aspect ratio $(length^{2}/area)$ as a function of time.}
\label{fig:strainless} 
\end{figure}

This computation yields 73 elliptic LCSs, computed as robust limit cycles of the differential equation family \eqref{eq:lambdaODE} (Fig. \ref{fig:closed_orbits_t0}). This set of closed curves forms a generalized KAM region, filled with material loops that resist filamentation and act as coherent transport barriers through the entire duration of the underlying video footage.

Figure \ref{fig:GRS_t0} shows separately the elliptic LCS with perfect
coherence ($\lambda=1$) in black, as well as the outermost elliptic LCS in blue that
forms the outer boundary of the coherent Lagrangian vortex associated with the GRS. This vortex boundary is marked by the parameter value $\lambda=1.0063$, which forecasts a roughly $0.6\%$ increase in arclength. The two closed curves enclose a highly coherent annular barrier, the Lagrangian counterpart of the outer ring identified within a collar of the GRS described in \cite{Liu}. This outer ring was constructed as the annulus outside the perceived core of the collar, a closed curve of velocity maxima.

Our geometric construction of an annular Lagrangian transport barrier is also motivated by the visual observations in \cite{Pater10}. These often indicate a sharper inner boundary and a more diffusive outer boundary for coherent jovian rings. The sharp observational boundary suggests an elliptic LCS of the highest possible coherence (zero stretching), while a diffusive outer boundary is expected near an elliptic LCS of the lowest possible coherence (highest stretching in a family of elliptic LCSs). 

The exact location of these oval barriers is expected to change under varying data resolution. The structural stability in the construction of elliptic LCSs, however, guarantees that the barriers move only by a small amount under small enough variations in resolution.

Advected images of the extracted GRS boundary confirm its sustained coherence over the finite time of extraction (see \Cref{fig:closed_orbits_T,fig:GRS_T,fig:length_change}). Our computation shows that the predicted coherent core of the GRS indeed regains its arc-length after 24 Jovian days. At the same time, the coherent outer boundary of the GRS indeed grows in arclength by about $0.6\%$, while its longitudinal extent decreases by about $5\%$. This suggests that the coherent boundary of the GRS is becoming rounder, which is generally consistent with the available observational records taken over much longer periods \cite{Simon02} (see Fig. \ref{fig:aspect_ratio}). Clearly, any firm conclusion or prediction about the long-term behavior of the arclength of material GRS boundary would require the analysis of a substantially longer data set.

\section{Summary}

We have applied the recently developed geodesic theory of transport
barriers \cite{black_hole,faraz13} to an enhanced video from NASA's
Cassini mission to Jupiter. First, we obtained a representative two-dimensional
wind field from this video via the Advection Corrected Correlation
Image Velocimetry (ACCIV) algorithm of Asay--Davis et al. \cite{ACCIV}.
Next, we identified, for the first time, unsteady material transport barriers
in the wind field that form the cores of zonal jets and the boundary
of the Great Red Spot (GRS) in Jupiter's atmosphere.

The parabolic LCSs (Lagrangian jet cores) we have found show that
both easterly and westerly jet cores provide strong material transport
barriers. This latter finding confirms the conclusion of Beron-Vera
et al. \cite{Beron_Vera08b} based on a numerical study of a perturbed
potential-vorticity-staircase model relevant for Jupiter. Deforming
material blobs placed near the parabolic LCS also reveal the Lagrangian
footprint of the recently discovered chevron-type atmospheric features
\cite{Simon12}.

The elliptic LCSs we identify provide a foliation of the GRS into highly coherent, uniformly stretching layers. This supports the existence of a proposed two-dimensional, cylindrical material transport barrier around the GRS \cite{Conrath81,Pater10}. According to our results, this cylindrical region has finite width, represented by an annulus in our two dimensional analysis. The annulus has a perfectly coherent ($\lambda=1$) inner boundary and a nearly perfectly coherent outer boundary, as shown in Fig. \ref{fig:GRS_t0}. While the outer boundary shrinks in longitudinal extent over the observed 24 Jovian days, its total arc-length shows a slight increase of about $0.6\%$. This suggests a modest evolution of this Lagrangian boundary towards more perfect circularity, which is in line with longer observational records \cite{Simon02}.

The time-resolved image reconstruction technique employed here is
purely kinematic, and does not incorporate a fit to dynamic equations
believed to govern Jupiter's wind fields. Considerable effort has
been devoted to fitting dynamically consistent reduced models to optically
reconstructed velocities (see, e.g., \cite{Shetty10}). These models,
however, are steady in a moving frame, and incorporate velocity measurements
from different sources and times. Here, instead of securing dynamical
consistency for an averaged, steady velocity field, we have constructed
an unsteady velocity field that is kinematically consistent with a
specific observational period. Imposing some degree of dynamical consistency on the optically reconstructed velocity field and comparing it with steady models in a moving frame remains a viable future research direction. A clear challenge is that the cloud distribution
over the GRS does not align with the location of its associated potential
vorticity anomaly or with any other of the GRS's known dynamical features
\cite{ACCIV}.

Arriving at Jupiter in 2016, the Juno mission of NASA will explore some of the material movement deep beneath the planet's clouds for the first time  \cite{NASA_link3}. Using this future information, we expect to be able to extend some aspects of our present analysis to three-dimensions using recently developed 3D variational LCS methods \cite{Blazevski14}.

We envision further applications of the methodology
developed here to remotely observed patterns in meteorology \cite{Peng},
oceanography \cite{Beron-Vera08}, environmental monitoring \cite{Wei,Mackenzie}
and crowd surveillance \cite{Ali,Surana}.

\section*{Acknowledgments}

We are thankful to Francisco Beron--Vera, David Choi, Andrew Ingersoll,
Adam Showman, Amy Simon-Miller, and Mohammad Farazmand for useful
interactions and for the related materials they shared with us. We would also like to acknowledge helpful discussions with Erik Bollt and Ranil Basnayake regarding another video footage of Jupiter. 

\pagebreak{}

\end{document}